\documentclass{aa}  

\usepackage{graphicx}

\usepackage[varg]{txfonts}

\usepackage[colorlinks=true, linkcolor=blue, citecolor=blue, urlcolor=blue]{hyperref}

\usepackage{tikz}
\usepackage{circuitikz}
\usetikzlibrary{positioning, shapes.geometric, decorations.markings, circuits.ee.IEC}
\usepackage{orcidlink}
\usepackage{amssymb}
\usepackage{gensymb}
\usepackage{textcomp}
\usepackage{appendix}   
\usepackage[none]{hyphenat}
\usepackage{float}
\usepackage{wasysym}
\usepackage{footnotehyper}
\usepackage[normalem]{ulem}
\usepackage{soul}
\usepackage{makecell}
\usepackage{bbm}
\usepackage{dsfont}

\begin{document} 

\sloppy 

   \title{Photonic integrated circuits for astronomy: A formal description of an integrated photonics-based wavefront sensor (IP-WFS)}
\titlerunning{A\&A, XXX, AXXX (2025)}

   \author{Diego Portero-Rodríguez
          \inst{1,2}\orcidlink{0000-0002-3191-5284},
          Hugo García-Vázquez\inst{1}\fnmsep\thanks{Corresponding author \email{hugo.garciavazquez@iac.es}}\orcidlink{0000-0003-0942-5761},
          José Javier Díaz García\inst{1}\orcidlink{0000-0003-2101-1078}, Luis Fernando Rodríguez Ramos\inst{1}\orcidlink{0000-0001-9968-0680},
          Félix Gracia Témich\inst{1}\orcidlink{0000-0002-4195-2337}
          \and
          J. Alfonso L. Aguerri\inst{1}\orcidlink{0000-0002-2839-2144}
          }

   \institute{Instituto de Astrofísica de Canarias (IAC), Área de Instrumentación, Departamento de Electrónica, \\Laboratorio de Circuitos Integrados (LABIC), 38205 La Laguna, Tenerife, Spain
         \and
             Universidad de La Laguna, Departamento de Astrofísica, 38206 La Laguna, Tenerife, Spain
             }

   \date{Received August 29, 2025; accepted April 20, 2026}
\authorrunning{Portero-Rodríguez, D., et al.}

  \abstract
    {Solar wavefront sensing has been a challenge for astrophysical instrumentalists, due to the low contrast between the Sun and the sky background compared to night-time observations, which limits the performance of adaptive optics systems.}
   {Wavefront correction in solar physics requires the analysis of extended images; meanwhile, at night  the displacement of a punctual object is analysed. This technique limits the spatial resolution, and therefore the accuracy in the wavefront reconstruction. }
   {To solve this problem, a new method of direct wavefront sensing without the need for image formation was explored for this work. A novel and promising technology called integrated photonics was used to accomplish this task. It allows the direct measurement of phase differences across the wavefront without the need to form images, using the principle of interferometry. This technology offers a low-consumption, miniaturised solution to astrophysical problems.}
   {For this work a mathematical model was derived to characterise the behaviour of the proposed wavefront sensor. The proposed system was verified and simulated using a Python-based adaptive optics simulator. These simulations demonstrate the physical behaviour of the proposed wavefront sensor and highlight the factors that must be taken into account for its correct functioning.}
   {}


   \maketitle
   \nolinenumbers

\section{Introduction}

Astrophotonics is a field of physics that employs the application of photonics to astronomical instruments. This results in the development of novel instruments and functionalities that were previously unfeasible, which in turn enables the reduction in size and consumption of optical instruments operating at visible and near-infrared wavelengths. This is achieved by replacing traditional optical elements with photonic devices that can manipulate light within waveguides \citep{ellis2024astrophotonicsrecentfuturedevelopments}. Some of these instruments are already in use;  one of the most successful is the photonic lantern developed by the University of Sydney \citep{Birks:15}. This device can be used to couple light from the focal plane of the telescope onto single-mode fibres, which has generally been avoided until now due to the complexity of light coupling compared with multimode fibres. This leads to direct applications in small satellite instrumentation and exoplanet research, and has been tested in the 3.9m Anglo-Australian Telescope (AAT) in Australia to suppress unwanted noise sources in astronomical observations \citep{LeonSavalArgyrosBlandHawthorn+2013+429+440}.

Within astrophotonics, there has been a noticeable increase in the development of a particular technology, known as photonic integrated circuits (PICs) or photonic chips \citep{Norris_2019,10.1117/12.2655630}. This technology has the potential to facilitate the miniaturisation and enhancement of astronomical instruments. Instruments such as coronagraphs for biosignature searches with the Habitable Worlds Observatory \citep{10.1117/12.3020518} or an integrated heterodyne polarimeter to measure the B-mode (or curl mode) polarisation of the cosmic microwave background (CMB; \citealt{inventions8060135}) are already under development. The first starlight spectrum obtained using an integrated spectrograph was reported by \citep{refId1}, who demonstrated that the simultaneous acquisition of multiple spectra using a single on-chip micro-spectrograph was feasible. Among all these applications, the beam combiner of the Gravity VLTI instrument \citep{GRAVITY, 10.1117/12.856689} is probably the most successful application  to  date of integrated photonics to astronomy.

Adaptive optics is another popular technique used in astronomy to correct wavefront aberrations, either in day or night observations. Classical adaptive optics has been widely used in solar telescopes. The enhancement of the spatial resolution of solar observations has been demonstrated by bringing telescopes closer to the diffraction limit. This helps us to understand the physics of small-scale structures of the Sun and to explain physical phenomena that remain unresolved \citep{Rimmele2011}. It is currently the most popular technique for wavefront aberration correction, and telescopes such as the GREGOR solar telescope \citep{GREGORAO} or the Swedish 1-m Solar Telescope \citep{refId0} employs it. Improved mechanisms such as multi-conjugate adaptive optics (MCAO) or ground-layer adaptive optics have gained popularity in recent years and have been employed in telescopes such as the Dune Solar Telescope (DST; \citealt{2008amos}) or the New Vacuum Solar Telescope (NVST) placed at the Fuxian Lake Solar Observatory \citep{zhanglanqiang}.

One of the key components that can be found in all the variations of adaptive optics are the wavefront sensors (WFSs). There are numerous examples of the application of WFSs in solar observations in the literature. These include the plenoptic camera proposed for the VTT solar telescope \citep{inproceedings}, the Shack-Hartmann WFS used in the THEMIS solar telescope \citep{tallon2021wavefrontsensingmakingofthemis}, and the curvature sensor proposed by \citep{article}. These systems are widely employed during day and night-time observations, although the references they use are different \citep{10.1117/12.2233528}. Due to the lack of naturally occurring guided stars with enough brightness to be useful in AO systems, laser guide stars (LGSs) are employed as a reference point for night-time observations \citep{refId01}. In day-time observations, AO systems become more challenging due to the extended nature of the Sun. It is therefore necessary to employ solar structures such as granulations, pores, or sunspots \citep{refId02} as a reference to compute phase aberrations. This limits the spatial resolution of the sensed wavefront and provides enough information to use only the actuators independently if they are separated by a distance greater than this limit.  Therefore, the performance of the AO system seems restricted by limiting the size of the microlenses used in the WFS to a minimum.

To address this issue, this group proposed a novel integrated photonics-based wavefront sensor in \citep{arrierolopez:hal-04414772}, continuing this work with a mathematical description and simulations of an ideal 2 pixel version \citep{10966318}. This work presents an improved version of the wavefront sensor that takes into account and analyses higher pixel densities and non-idealities. In addition, a formal description of the working principle of this astrophotonic device is presented, along with an analysis of potential enhancements that could be achieved by using this wavefront sensor. 

The paper is structured as follows. Section \ref{Wavefront sensor analysis} presents a proposal for an adaptive optic system and a more detailed analysis of the working principle and design of the WFS, together with a mathematical description. Section \ref{Integrated photonic chip development platforms} presents the different integrated photonic chip development platforms that are currently available, and their advantages and disadvantages. Section \ref{Light coupling} discusses the issue of light coupling to photonic devices and, in particular, to the proposed wavefront sensor. Section \ref{Simulation_method} presents the pipeline followed for implementing a full system simulation of the adaptive optics system and of the IP-WFS. Section \ref{results} presents the results obtained in the full system simulation. Section \ref{Summary} presents the conclusions obtained from this work.

\section{Wavefront sensor analysis}

\label{Wavefront sensor analysis}

In this section an analysis of the working principle of the proposed wavefront sensor and a mathematical description of the photonic integrated circuit are presented. The analysis was conducted for telescope sizes up to 4m class. For solar telescopes, this represents the largest solar telescopes currently in use, such as the \textit{Daniel K. Inouye} Solar Telescope (DKIST; \citealt{Rimmele2020_DKIST}) at the Haleakala Observatory in Hawaii and the future European Solar Telescope (EST; \citealt{refId0_EST})  at the Roque de los Muchachos Observatory in La Palma, Spain.

\begin{figure}[h]
\centering
\includegraphics[scale=0.063]{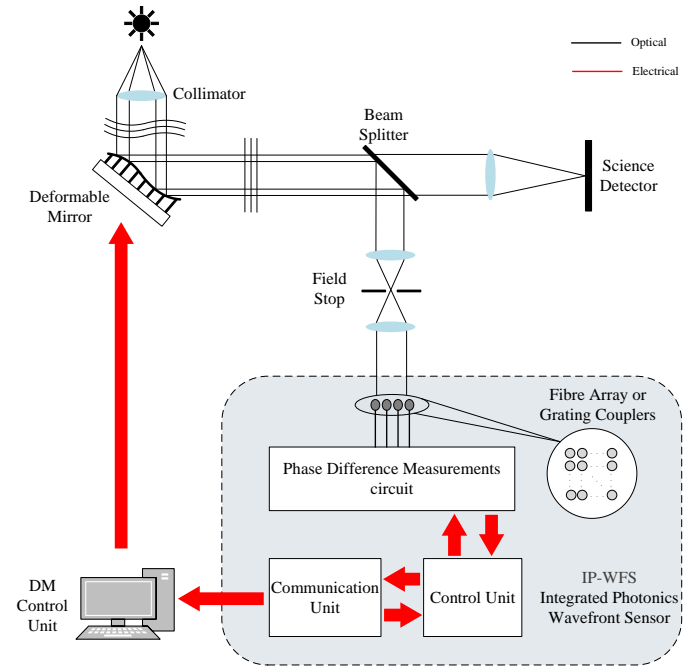}
\caption{Adaptive optic system with IP-WFS. Different regions of the wavefront are coupled to the phase difference measurement circuit using a fibre array or grating coupler. A deformable mirror then corrects the wavefront aberrations using the information provided by the wavefront sensor.}
          \label{Adaptive-optic-system}
\end{figure}

The proposed wavefront sensor should be able to detect wavefront aberrations without the need to form images, based on the phenomenon of interferometry. Figure \ref{Adaptive-optic-system} depicts the proposed adaptive optic system. In this configuration, the light emitted by the Sun is collimated, corrected by the deformable mirror, and split in a certain ratio, with some of the light directed to the science detector and the rest to a field stop and subsequently to the wavefront sensor. The light is then coupled to the integrated circuit (IC), which calculates the phase shifts existing in the wavefront. The result is then transmitted as an electrical signal to a control unit, which generates the signals necessary to control the different components of the wavefront sensor. Through a communication unit, the system will communicate with the wavefront sensor and the deformable mirror (DM) control unit to actuate over the DM.

\subsection{System description}

The light is coupled to the wavefront sensor through a microlens array. There are then two main approaches to light coupling: using optical fibres with edge couplers, or using grating couplers. The issue of light coupling is discussed more deeply in Sect. \ref{Light coupling}. The wavefront sensor computes phase differences across the wavefront using the phenomenon of interferometry. This interferometric result is conveyed as an electrical signal to a signal conditioning stage that digitises the readout signals and generates the control signals required for controlling the photonic components used within the integrated circuit.

\begin{figure}[h]
   \centering
   \resizebox{0.44\textwidth}{!}{
      \includegraphics{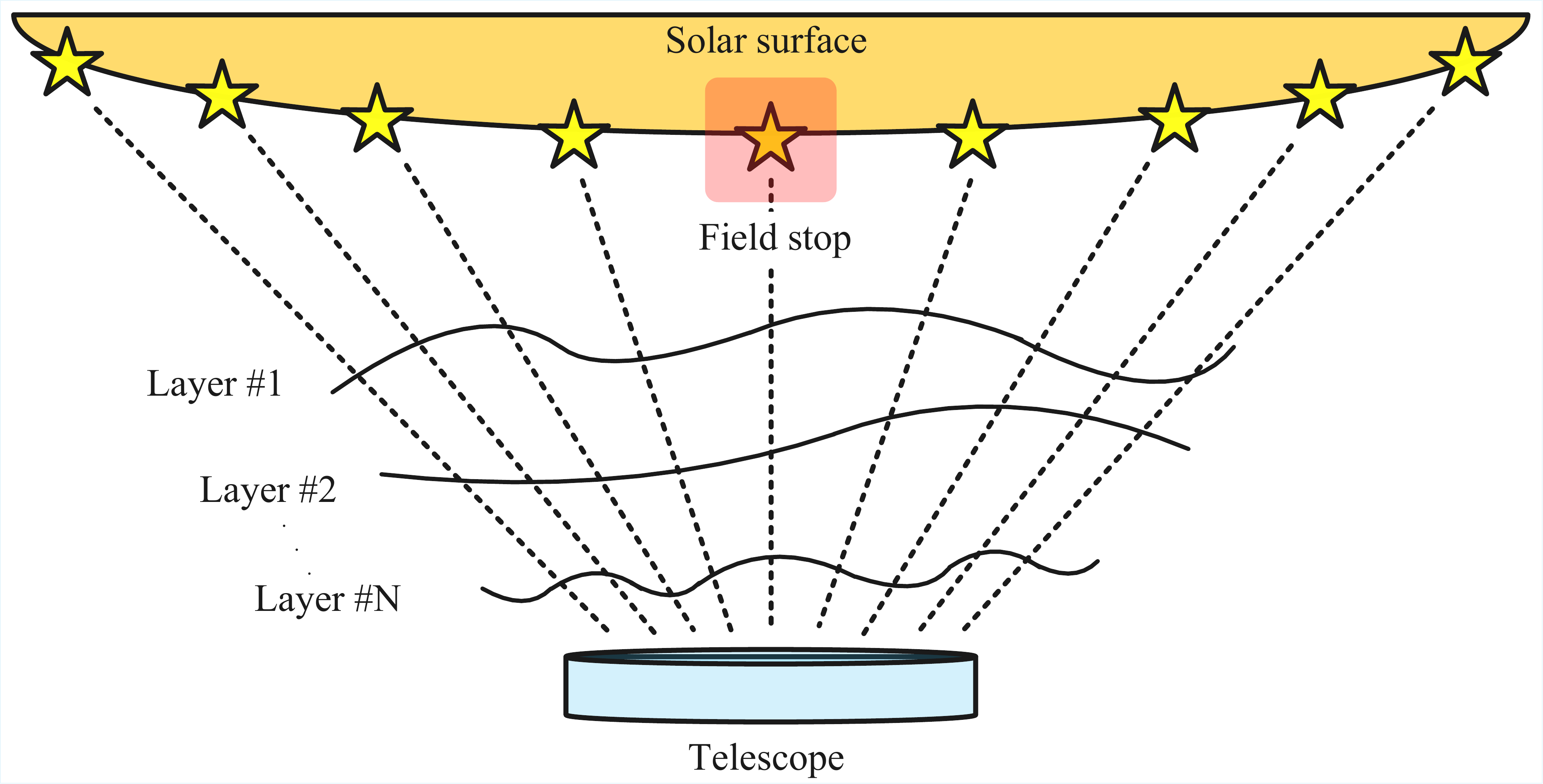}}
   \caption{Depending on the angle of view, different portions of the Sun undergo different optical paths. This phenomenon is known as anisoplanatic error. A field stop is used to restrict the field of view, thereby ensuring that the wavefront can be approximated as   a point source. The layers represent atmospheric layers at different heights.}
   \label{Field-stop}
\end{figure}

One problem that arises in solar adaptive optics and to which this wavefront sensor is especially sensitive, is the anisoplanatic error. This error occurs because different points in an extended source undergo different optical paths depending on their angle relative to a reference direction \citep{Bos:24}. In practice, this means that different parts of the Sun will reach the telescope with different wavefronts, and the resulting pupil image will be a combination of all of them. To solve this problem, a field stop is placed in the focal plane and wavefront sensing is performed using a small slice of the field (see Fig. \ref{Field-stop}). The field stop should have a minimum size to ensure that all the points in its wavefront pass through the same atmospheric volume. If this happens, the wavefront can be  considered approximately the same as for a point source. This has the same effect as single conjugate adaptive optics (SCAO), where wavefront corrections deviating from the line of sight will experience an anisoplanatic error.

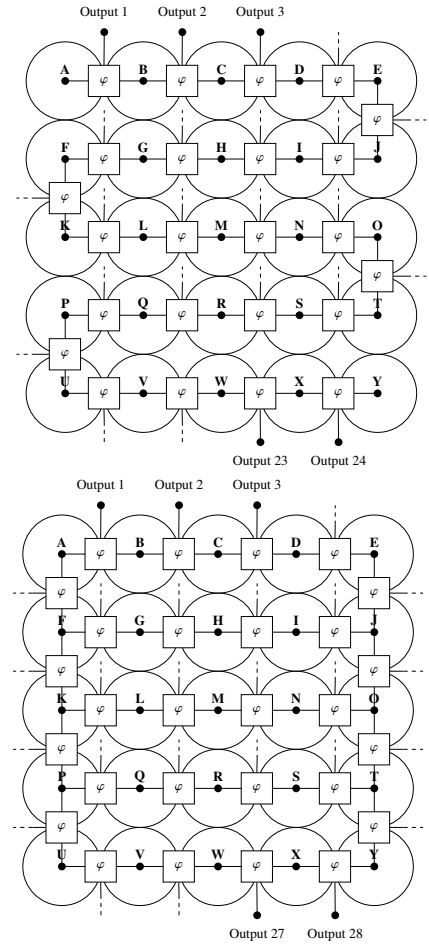
\begin{figure}[h]
    \centering
    \resizebox{0.3\textwidth}{!}{
        \begin{circuitikz}

    \foreach \row in {0,1,2,3,4} {
        \foreach \col in {0,1,2,3,4} {
            \pgfmathtruncatemacro{\letter}{\row * 5 + \col + 65} 
            \draw (\col * 2, -\row * 2) circle(1); 
            \node at (\col * 2, -\row * 2 + 0.3) {\textbf{\char\letter}}; 
            \fill (\col * 2, -\row * 2) circle(0.1); 
        }
    }

    \draw (0.4, -3) node[draw, minimum size=0.8cm, anchor=east, fill=white] (Phi1){$\varphi$};

    \draw (0.4, -7) node[draw, minimum size=0.8cm, anchor=east, fill=white](Phi2) {$\varphi$} ;

    \draw (8.4, -1) node[draw, minimum size=0.8cm, anchor=east, fill=white](Phi3) {$\varphi$} ;

    \draw (8.4, -5) node[draw, minimum size=0.8cm, anchor=east, fill=white] (Phi4){$\varphi$} ;
    
    \draw (1.4, 0) node[draw, minimum size=0.8cm, anchor=east, fill=white] (Phi5){$\varphi$};

    \draw (3.4, 0) node[draw, minimum size=0.8cm, anchor=east, fill=white] (Phi6){$\varphi$};

    \draw (5.4, 0) node[draw, minimum size=0.8cm, anchor=east, fill=white](Phi7) {$\varphi$};

    \draw (7.4, 0) node[draw, minimum size=0.8cm, anchor=east, fill=white] (Phi8){$\varphi$};
    
    \draw (1.4, -2) node[draw, minimum size=0.8cm, anchor=east, fill=white](Phi9) {$\varphi$};

    \draw (3.4, -2) node[draw, minimum size=0.8cm, anchor=east, fill=white](Phi10) {$\varphi$};

    \draw (5.4, -2) node[draw, minimum size=0.8cm, anchor=east, fill=white](Phi11) {$\varphi$};

    \draw (7.4, -2) node[draw, minimum size=0.8cm, anchor=east, fill=white](Phi12) {$\varphi$};

    \draw (1.4, -4) node[draw, minimum size=0.8cm, anchor=east, fill=white] (Phi13){$\varphi$};

    \draw (3.4, -4) node[draw, minimum size=0.8cm, anchor=east, fill=white](Phi14) {$\varphi$};

    \draw (5.4, -4) node[draw, minimum size=0.8cm, anchor=east, fill=white] (Phi15){$\varphi$};

    \draw (7.4, -4) node[draw, minimum size=0.8cm, anchor=east, fill=white] (Phi16){$\varphi$};

    \draw (1.4, -6) node[draw, minimum size=0.8cm, anchor=east, fill=white] (Phi17){$\varphi$};

    \draw (3.4, -6) node[draw, minimum size=0.8cm, anchor=east, fill=white] (Phi18){$\varphi$};

    \draw (5.4, -6) node[draw, minimum size=0.8cm, anchor=east, fill=white] (Phi19){$\varphi$};

    \draw (7.4, -6) node[draw, minimum size=0.8cm, anchor=east, fill=white] (Phi20){$\varphi$};

    \draw (1.4, -8) node[draw, minimum size=0.8cm, anchor=east, fill=white] (Phi21){$\varphi$};

    \draw (3.4, -8) node[draw, minimum size=0.8cm, anchor=east, fill=white] (Phi22){$\varphi$};

    \draw (5.4, -8) node[draw, minimum size=0.8cm, anchor=east, fill=white] (Phi23){$\varphi$};

    \draw (7.4, -8) node[draw, minimum size=0.8cm, anchor=east, fill=white] (Phi24){$\varphi$};

    \draw (0, 0) -- (0.6, 0);

    \draw (1.4, 0) -- (2.6, 0);

    \draw (3.4, 0) -- (4.6, 0);

    \draw (5.4, 0) -- (6.6, 0);

    \draw (7.4, 0) -- (8, 0);

    \draw (8, 0) -- (8, -0.6);

    \draw (8, -1.4) -- (8, -2);

    \draw (8, -2) -- (7.4, -2);

    \draw (6.6, -2) -- (5.4, -2);

    \draw (4.6, -2) -- (3.4, -2);

    \draw (2.6, -2) -- (1.4, -2);

    \draw (0.6, -2) -- (0, -2);

    \draw (0, -2) -- (0, -2.6);

    \draw (0, -3.4) -- (0, -4);

    \draw (0, -4) -- (0.6, -4);

    \draw (2.6, -4) -- (1.4, -4);

    \draw (4.6, -4) -- (3.4, -4);

    \draw (6.6, -4) -- (5.4, -4);

    \draw (0, -6) -- (0.6, -6);

    \draw (2.6, -6) -- (1.4, -6);

    \draw (4.6, -6) -- (3.4, -6);

    \draw (6.6, -6) -- (5.4, -6);

    \draw (0, -8) -- (0.6, -8);

    \draw (2.6, -8) -- (1.4, -8);

    \draw (4.6, -8) -- (3.4, -8);

    \draw (6.6, -8) -- (5.4, -8);

    \draw (7.4, -8) -- (8, -8);

    \draw (8, -6) -- (8, -5.4);

    \draw (8, -6) -- (7.4, -6);

    \draw (8, -4) -- (8, -4.6);

    \draw (8, -4) -- (7.4, -4);

    \draw (0, -6) -- (0, -6.6);

    \draw (0, -7.4) -- (0, -8);

    \draw[] (Phi1.west) -- (-0.75, -3);

    \draw[dashed] (-0.75, -3) -- (-1.25, -3);

    \draw[] (Phi2.west) -- (-0.75, -7);

    \draw[dashed] (-0.75, -7) -- (-1.25, -7);

    \draw[] (Phi3.east) -- (8.75, -1);

    \draw[dashed] (8.75, -1) -- (9.25, -1);

    \draw[] (Phi4.east) -- (8.75, -5);

    \draw[dashed] (8.75, -5) -- (9.25, -5);

    \draw[] (Phi5.north) -- (1, 1.25);
    
    \fill (1, 1.25) circle(0.1); 
    
    \node at (1, 1.75) {Output 1};

    \draw[] (Phi6.north) -- (3, 1.25);
    \fill (3, 1.25) circle(0.1);
    \node at (3, 1.75) {Output 2};

    \draw[] (Phi7.north) -- (5, 1.25);
    \fill (5, 1.25) circle(0.1);
    \node at (5, 1.75) {Output 3};

    \draw[] (Phi8.north) -- (7, 0.75);

    \draw[dashed] (7, 0.75) -- (7, 1.25);

    \draw[] (Phi9.north) -- (1, -1.25);

    \draw[dashed] (1, -1.25) -- (1, -0.75);

    \draw[] (Phi10.north) -- (3, -1.25);

    \draw[dashed] (3, -1.25) -- (3, -0.75);

    \draw[] (Phi11.north) -- (5, -1.25);

    \draw[dashed] (5, -1.25) -- (5, -0.75);

    \draw[] (Phi12.north) -- (7, -1.25);

    \draw[dashed] (7, -1.25) -- (7, -0.75);

    \draw[] (Phi13.north) -- (1, -3.25);

    \draw[dashed] (1, -3.25) -- (1, -2.75);

    \draw[] (Phi14.north) -- (3, -3.25);

    \draw[dashed] (3, -3.25) -- (3, -2.75);

    \draw[] (Phi15.north) -- (5, -3.25);

    \draw[dashed] (5, -3.25) -- (5, -2.75);

    \draw[] (Phi16.north) -- (7, -3.25);

    \draw[dashed] (7, -3.25) -- (7, -2.75);

    \draw[] (Phi17.north) -- (1, -5.25);

    \draw[dashed] (1, -5.25) -- (1, -4.75);

    \draw[] (Phi18.north) -- (3, -5.25);

    \draw[dashed] (3, -5.25) -- (3, -4.75);

    \draw[] (Phi19.north) -- (5, -5.25);

    \draw[dashed] (5, -5.25) -- (5, -4.75);

    \draw[] (Phi20.north) -- (7, -5.25);

    \draw[dashed] (7, -5.25) -- (7, -4.75);

    \draw[] (Phi21.south) -- (1, -8.75);

    \draw[dashed] (1, -9.25) -- (1, -8.75);

    \draw[] (Phi22.south) -- (3, -8.75);

    \draw[dashed] (3, -9.25) -- (3, -8.75);

    \draw[] (Phi23.south) -- (5, -9.25);
    \fill (5, -9.25) circle(0.1);
    \node at (5, -9.75) {Output 23};

    \draw[] (Phi24.south) -- (7, -9.25);
    \fill (7, -9.25) circle(0.1);
    \node at (7, -9.75) {Output 24};

\end{circuitikz}
    }

    \resizebox{0.3\textwidth}{!}{
        \begin{circuitikz}

      \foreach \row in {0,1,2,3,4} {
        \foreach \col in {0,1,2,3,4} {
            \pgfmathtruncatemacro{\letter}{\row * 5 + \col + 65}
            \draw (\col * 2, -\row * 2) circle(1);
            \node at (\col * 2, -\row * 2 + 0.3) {\textbf{\char\letter}};  
            \fill (\col * 2, -\row * 2) circle(0.1); 
        }
    }

    \draw (0.4, -1) node[draw, minimum size=0.8cm, anchor=east, fill=white] (Phi25){$\varphi$} ;

    \draw (0.4, -5) node[draw, minimum size=0.8cm, anchor=east, fill=white] (Phi26){$\varphi$} ;

    \draw (0.4, -3) node[draw, minimum size=0.8cm, anchor=east, fill=white] (Phi1){$\varphi$} ;

    \draw (0.4, -7) node[draw, minimum size=0.8cm, anchor=east, fill=white](Phi2) {$\varphi$} ;

    \draw (8.4, -1) node[draw, minimum size=0.8cm, anchor=east, fill=white](Phi3) {$\varphi$} ;

    \draw (8.4, -5) node[draw, minimum size=0.8cm, anchor=east, fill=white] (Phi4){$\varphi$} ;

    \draw (8.4, -3) node[draw, minimum size=0.8cm, anchor=east, fill=white] (Phi27){$\varphi$} ;

    \draw (8.4, -7) node[draw, minimum size=0.8cm, anchor=east, fill=white] (Phi28){$\varphi$} ;

    \draw (1.4, 0) node[draw, minimum size=0.8cm, anchor=east, fill=white] (Phi5){$\varphi$};

    \draw (3.4, 0) node[draw, minimum size=0.8cm, anchor=east, fill=white] (Phi6){$\varphi$};

    \draw (5.4, 0) node[draw, minimum size=0.8cm, anchor=east, fill=white](Phi7) {$\varphi$};

    \draw (7.4, 0) node[draw, minimum size=0.8cm, anchor=east, fill=white] (Phi8){$\varphi$};
    
    \draw (1.4, -2) node[draw, minimum size=0.8cm, anchor=east, fill=white](Phi9) {$\varphi$};

     \draw (3.4, -2) node[draw, minimum size=0.8cm, anchor=east, fill=white](Phi10) {$\varphi$};

    \draw (5.4, -2) node[draw, minimum size=0.8cm, anchor=east, fill=white](Phi11) {$\varphi$};

    \draw (7.4, -2) node[draw, minimum size=0.8cm, anchor=east, fill=white](Phi12) {$\varphi$};

    \draw (1.4, -4) node[draw, minimum size=0.8cm, anchor=east, fill=white] (Phi13){$\varphi$};

    \draw (3.4, -4) node[draw, minimum size=0.8cm, anchor=east, fill=white](Phi14) {$\varphi$};

    \draw (5.4, -4) node[draw, minimum size=0.8cm, anchor=east, fill=white] (Phi15){$\varphi$};

    \draw (7.4, -4) node[draw, minimum size=0.8cm, anchor=east, fill=white] (Phi16){$\varphi$};

    \draw (1.4, -6) node[draw, minimum size=0.8cm, anchor=east, fill=white] (Phi17){$\varphi$};

    \draw (3.4, -6) node[draw, minimum size=0.8cm, anchor=east, fill=white] (Phi18){$\varphi$};

    \draw (5.4, -6) node[draw, minimum size=0.8cm, anchor=east, fill=white] (Phi19){$\varphi$};

    \draw (7.4, -6) node[draw, minimum size=0.8cm, anchor=east, fill=white] (Phi20){$\varphi$};

    \draw (1.4, -8) node[draw, minimum size=0.8cm, anchor=east, fill=white] (Phi21){$\varphi$};

    \draw (3.4, -8) node[draw, minimum size=0.8cm, anchor=east, fill=white] (Phi22){$\varphi$};

    \draw (5.4, -8) node[draw, minimum size=0.8cm, anchor=east, fill=white] (Phi23){$\varphi$};

    \draw (7.4, -8) node[draw, minimum size=0.8cm, anchor=east, fill=white] (Phi24){$\varphi$};   

    \draw (0, 0) -- (0.6, 0);

    \draw (0, 0) -- (0, -0.6);

    \draw (1.4, 0) -- (2.6, 0);

    \draw (3.4, 0) -- (4.6, 0);

    \draw (5.4, 0) -- (6.6, 0);

    \draw (7.4, 0) -- (8, 0);

    \draw (8, 0) -- (8, -0.6);

    \draw (8, -1.4) -- (8, -2);

    \draw (8, -2) -- (7.4, -2);

    \draw (6.6, -2) -- (5.4, -2);

    \draw (4.6, -2) -- (3.4, -2);

    \draw (2.6, -2) -- (1.4, -2);

    \draw (0.6, -2) -- (0, -2);

    \draw (0, -2) -- (0, -2.6);

    \draw (0, -2) -- (0, -1.4);

    \draw (0, -3.4) -- (0, -4);

    \draw (0, -4) -- (0.6, -4);

    \draw (0, -4) -- (0, -4.6);

    \draw (0, -5.4) -- (0, -6);

    \draw (8, -3.4) -- (8, -4);

    \draw (8, -2.6) -- (8, -2);

    \draw (8, -8) -- (8, -7.4);

    \draw (8, -6.6) -- (8, -6);

    \draw (2.6, -4) -- (1.4, -4);

    \draw (4.6, -4) -- (3.4, -4);

    \draw (6.6, -4) -- (5.4, -4);

    \draw (0, -6) -- (0.6, -6);

    \draw (2.6, -6) -- (1.4, -6);

    \draw (4.6, -6) -- (3.4, -6);

    \draw (6.6, -6) -- (5.4, -6);

    \draw (0, -8) -- (0.6, -8);

    \draw (2.6, -8) -- (1.4, -8);

    \draw (4.6, -8) -- (3.4, -8);

    \draw (6.6, -8) -- (5.4, -8);

    \draw (7.4, -8) -- (8, -8);

    \draw (8, -6) -- (8, -5.4);

    \draw (8, -6) -- (7.4, -6);

    \draw (8, -4) -- (8, -4.6);

    \draw (8, -4) -- (7.4, -4);

    \draw (0, -6) -- (0, -6.6);

    \draw (0, -7.4) -- (0, -8);

    \draw[] (Phi1.west) -- (-0.75, -3);

    \draw[dashed] (-0.75, -3) -- (-1.25, -3);

    \draw[] (Phi25.west) -- (-0.75, -1);

    \draw[dashed] (-0.75, -1) -- (-1.25, -1);

    \draw[] (Phi2.west) -- (-0.75, -7);

    \draw[dashed] (-0.75, -7) -- (-1.25, -7);

    \draw[] (Phi26.west) -- (-0.75, -5);

    \draw[dashed] (-0.75, -5) -- (-1.25, -5);

    \draw[] (Phi3.east) -- (8.75, -1);

    \draw[dashed] (8.75, -1) -- (9.25, -1);

    \draw[] (Phi27.east) -- (8.75, -3);

    \draw[dashed] (8.75, -3) -- (9.25, -3);

    \draw[] (Phi4.east) -- (8.75, -5);

    \draw[dashed] (8.75, -5) -- (9.25, -5);

    \draw[] (Phi28.east) -- (8.75, -7);

    \draw[dashed] (8.75, -7) -- (9.25, -7);

    \draw[] (Phi5.north) -- (1, 1.25);
    \fill (1, 1.25) circle(0.1);
    \node at (1, 1.75) {Output 1};

    \draw[] (Phi6.north) -- (3, 1.25);
    \fill (3, 1.25) circle(0.1);
    \node at (3, 1.75) {Output 2};

    \draw[] (Phi7.north) -- (5, 1.25);
    \fill (5, 1.25) circle(0.1);
    \node at (5, 1.75) {Output 3};

    \draw[] (Phi8.north) -- (7, 0.75);

    \draw[dashed] (7, 0.75) -- (7, 1.25);

    \draw[] (Phi9.north) -- (1, -1.25);

    \draw[dashed] (1, -1.25) -- (1, -0.75);

    \draw[] (Phi10.north) -- (3, -1.25);

    \draw[dashed] (3, -1.25) -- (3, -0.75);

    \draw[] (Phi11.north) -- (5, -1.25);

    \draw[dashed] (5, -1.25) -- (5, -0.75);

    \draw[] (Phi12.north) -- (7, -1.25);

    \draw[dashed] (7, -1.25) -- (7, -0.75);

    \draw[] (Phi13.north) -- (1, -3.25);

    \draw[dashed] (1, -3.25) -- (1, -2.75);

    \draw[] (Phi14.north) -- (3, -3.25);

    \draw[dashed] (3, -3.25) -- (3, -2.75);

    \draw[] (Phi15.north) -- (5, -3.25);

    \draw[dashed] (5, -3.25) -- (5, -2.75);

    \draw[] (Phi16.north) -- (7, -3.25);

    \draw[dashed] (7, -3.25) -- (7, -2.75);

    \draw[] (Phi17.north) -- (1, -5.25);

    \draw[dashed] (1, -5.25) -- (1, -4.75);

    \draw[] (Phi18.north) -- (3, -5.25);

    \draw[dashed] (3, -5.25) -- (3, -4.75);

    \draw[] (Phi19.north) -- (5, -5.25);

    \draw[dashed] (5, -5.25) -- (5, -4.75);

    \draw[] (Phi20.north) -- (7, -5.25);

    \draw[dashed] (7, -5.25) -- (7, -4.75);

    \draw[] (Phi21.south) -- (1, -8.75);

    \draw[dashed] (1, -9.25) -- (1, -8.75);

    \draw[] (Phi22.south) -- (3, -8.75);

    \draw[dashed] (3, -9.25) -- (3, -8.75);

    \draw[] (Phi23.south) -- (5, -9.25);
    \fill (5, -9.25) circle(0.1);
    \node at (5, -9.75) {Output 27};

    \draw[] (Phi24.south) -- (7, -9.25);
    \fill (7, -9.25) circle(0.1);
    \node at (7, -9.75) {Output 28};

\end{circuitikz}
    }
    \caption{Phase difference measurement potential configurations. The top panel shows the first configuration, which corresponds to a snake pattern. The bottom panel shows a more robust configuration that is more expensive.}
    \label{fig:phase_measurements}
\end{figure}

The primary objective of the photonic circuit is to calculate phase shifts within the wavefront. Figure \ref{fig:phase_measurements} shows two potential configurations for implementing the wavefront sensor. These configurations are arranged in a 5x5 array for ease of explanation, where each optical fibre is represented by a letter ranging from A to Y. In practice, more extensive arrays could be implemented, improving the resolution of the wavefront sensor. With the first configuration, the wavefront sensor will measure phase differences between the different inputs; therefore, its output is a relative measurement. To obtain a reconstruction of the wavefront, a reference must be defined. The main problem with this configuration arises if any of these comparison blocks is damaged. For example, if the block responsible for comparing fibres J and I were to fail, J would lose A as a reference, leading to two divided groups with different references and no relationship between them.

The second configuration represents a more robust configuration against systematic errors than the first. In the event of a block failure, the reference does not get lost, and continued functionality is enabled. Although it uses a larger number of components, it ensures that each optical fibre has a different possible optical path to the reference.

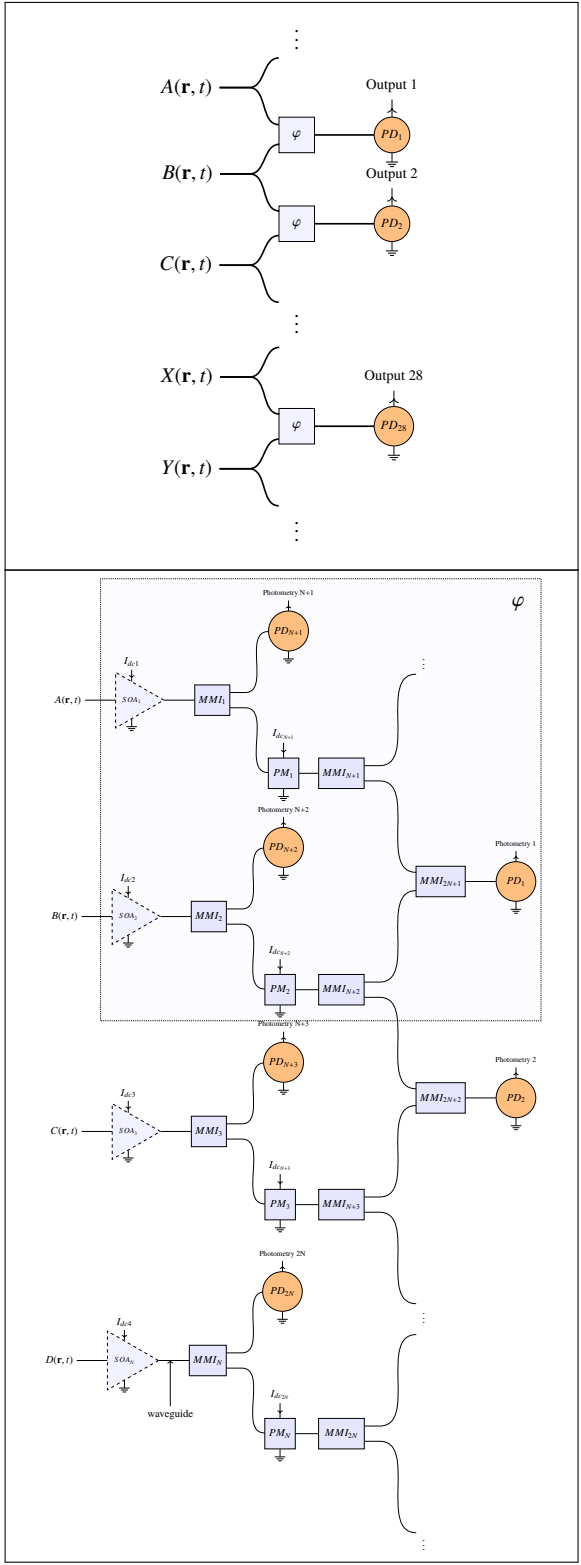
\begin{figure}[h]
    \centering
    \fbox{
        \begin{minipage}{0.39\textwidth}
            \centering
            \resizebox{0.53\textwidth}{!}{
                    \begin{tikzpicture}[
                circuit ee IEC,
                node distance=1.2cm and 1cm,
                block/.style={draw, minimum size=1cm, fill=blue!5},
                photodiode/.style={draw, circle, fill=orange!50, minimum width=0.5cm, minimum height=1cm},
                waveguide/.style={draw, thick, -},
                amplifier/.style={draw, regular polygon, regular polygon sides=3, align=center},
                decoration = {markings}
                ]
    
                \node[] (SOA1) {};
    
                \node[] (SOA2) {};

                \node[left=of SOA1, yshift=1.45cm, xshift=-0.18cm,font=\fontsize{10}{10}\selectfont\bfseries] (input1) {$Y(\mathbf{r},t)$};

                \node[block, scale=0.6, above=of SOA1, yshift=0.9cm, font=\fontsize{12}{10}\selectfont] (SOA5) {$\varphi$};
    
                \node[block, scale=0.6, above=of SOA1, yshift=6.6cm, font=\fontsize{12}{10}\selectfont] (SOA6) {$\varphi$};
    
                \node[block, scale=0.6, above=of SOA1, yshift=9.1cm, font=\fontsize{12}{10}\selectfont] (SOA7) {$\varphi$};
    
                \node[left=of SOA2, yshift=6.43cm, xshift=-0.17cm, font=\fontsize{10}{10}\selectfont\bfseries](input5) {$B(\mathbf{r},t)$};
    
                \node[left=of SOA6, yshift=-0.7cm, xshift=-0cm,font=\fontsize{10}{10}\selectfont\bfseries] (input4) {$C(\mathbf{r},t)$};
                 
                \node[left=of SOA2, yshift=3cm, xshift=-0.17cm, font=\fontsize{10}{10}\selectfont\bfseries](input2) {$X(\mathbf{r},t)$};
    
                \node[left=of SOA7, yshift=0.8cm, font=\fontsize{10}{10}\selectfont\bfseries](input6) {$A(\mathbf{r},t)$};

                \node[photodiode,  scale=0.6, right=of SOA7] (PD2) {$PD_{1}$};
    
                \node[photodiode,  scale=0.6, right=of SOA5] (PD1) {$PD_{28}$}; 
    
                \node[photodiode,  scale=0.6, right=of SOA6] (PD4) {$PD_{2}$};

                \draw[waveguide] (input2.east) ++(0,0) -- ++(0.5,0) to[out=0, in=180] ++(0.5,0.5);
                \draw[waveguide] (input2.east)++(0,0) -- ++(0.5,0) to[out=0, in=180] ++(0.5,-1.125+0.5);
    
                \draw[waveguide] (input1.east) ++(0,0) -- ++(0.5,0) to[out=0, in=180] ++(0.5,0.5);
                \draw[waveguide] (input1.east)++(0,0) -- ++(0.5,0) to[out=0, in=180] ++(0.5,-1.125+0.5);

                \draw[waveguide] (input4.east) ++(0,0) -- ++(0.5,0) to[out=0, in=180] ++(0.5,0.5);
                \draw[waveguide] (input4.east)++(0,0) -- ++(0.5,0) to[out=0, in=180] ++(0.5,-1.125+0.5);
    
                \draw[waveguide] (input5.east) ++(0,0) -- ++(0.5,0) to[out=0, in=180] ++(0.5,0.5);
                \draw[waveguide] (input5.east)++(0,0) -- ++(0.5,0) to[out=0, in=180] ++(0.5,-1.125+0.5);
    
                \draw[waveguide] (input6.east) ++(0,0) -- ++(0.5,0) to[out=0, in=180] ++(0.5,0.5);
                \draw[waveguide] (input6.east)++(0,0) -- ++(0.5,0) to[out=0, in=180] ++(0.5,-1.125+0.5);

                \draw[waveguide] (PD2) -- (SOA7);
    
                \draw[waveguide] (PD4) -- (SOA6);

                \draw[waveguide] (PD1) -- (SOA5);
    
                \node[draw=none] (ellipsis1) at (0, 4) {$\vdots$};
    
                \node[draw=none] (ellipsis1) at (0, 8.8) {$\vdots$};
    
                \node[draw=none] (ellipsis1) at (0, 0.5) {$\vdots$};
    
                \draw [waveguide, scale=0.6, line width=0.25pt] (PD1.south) -- ++(0,-0.25) node[ground, below, rotate=270, scale=0.7, xshift=0.1cm, yshift = 0.25cm] {};
                \draw [waveguide, scale=0.6, line width=0.25pt] (PD2.south) -- ++(0,-0.25) node[ground, below, rotate=270, scale=0.7, xshift=0.1cm, yshift = 0.25cm] {};
                \draw [waveguide, scale=0.6, line width=0.25pt] (PD4.south) -- ++(0,-0.25) node[ground, below, rotate=270, scale=0.7, xshift=0.1cm, yshift = 0.25cm] {};

                \draw[waveguide, scale=0.6, postaction={decorate, decoration={markings, mark=at position 0.5 with {\arrow{>}}}}, line width=0.25pt] (PD2) -- ++(0,1) node[above, font=\fontsize{7}{10}\selectfont] {Output 1};
    
                \draw[waveguide, scale=0.6, postaction={decorate, decoration={markings, mark=at position 0.5 with {\arrow{>}}}}, line width=0.25pt] (PD1) -- ++(0,1) node[above, font=\fontsize{7}{10}\selectfont] {Output 28};
    
                \draw[waveguide, scale=0.6, postaction={decorate, decoration={markings, mark=at position 0.5 with {\arrow{>}}}}, line width=0.25pt] (PD4) -- ++(0,1) node[above, font=\fontsize{7}{10}\selectfont] {Output 2};

            \end{tikzpicture}
            }
        \end{minipage}
    }

    \vspace{-0.14em}

    \fbox{
        \begin{minipage}{0.39\textwidth}
            \centering
            \resizebox{0.93\textwidth}{!}{
                \begin{tikzpicture}[
                    circuit ee IEC,
                    node distance=4cm and 1cm,
                    block/.style={draw, minimum size=1cm, fill=blue!10},
                    block2/.style={draw, minimum size=1cm, minimum width=1.5cm, fill=blue!10},
                    block_dashed/.style={draw, dashed, minimum size=1cm, fill=blue!10},
                    photodiode/.style={draw, circle, fill=orange!50, minimum width=0.5cm, minimum height=1cm},
                    photodiode2/.style={draw, circle, fill=orange!50, minimum width=0.7cm, minimum height=1.3cm},
                    waveguide/.style={draw, thick, -},
                    amplifier/.style={draw, regular polygon, regular polygon sides=3, align=center, fill=blue!5},
                    decoration = {markings}
                    ]
        
                    \draw[dash pattern=on 1pt off 1pt, fill=blue!1] (-5.7,-12.8) rectangle (8.75,1.7);
        
                    \node[font=\fontsize{20}{20}\selectfont\bfseries] at (8, 0.9) {$\varphi$};
                
                    \node[amplifier, shape border rotate=-90, scale=0.6, dashed, above=of SOA1, font=\fontsize{12}{10}\selectfont, yshift=-11.7cm, xshift=-7.8cm] (SOA1) {$SOA_{1}$};
                    \node[left=of SOA1, font=\fontsize{10}{10}\selectfont\bfseries] (input1) {$A(\mathbf{r},t)$};
        
                    \node[block, right=of SOA1] (MMI1) {$MMI_{1}$};
        
                    \node[block, right=of MMI1, yshift=-2.4cm, xshift=0.25cm] (PM1) {$PM_{1}$};
        
                    \node[photodiode, above=of MMI1, yshift=-2.9cm, xshift=2.5cm] (PD4) {$PD_{N+1}$};
        
                    \node[](PD5){};
        
                    \node[block, below=of PD5, xshift=2.2cm, yshift=-0.05cm] (MMI5) {$MMI_{N+1}$};
        
                    \draw[waveguide] (MMI1.east) ++(0,0.25) -- ++(0.5,0) to[out=0, in=180] ++(0.75,2);
                    
                    \draw[waveguide] (MMI1.east)++(0,-0.25) -- ++(0.5,0) to[out=0, in=180] ++(0.75,-2.125);

                    \draw[waveguide] (input1) -- (SOA1);
                    \draw[waveguide] (SOA1) -- (MMI1);
        
                    \draw[waveguide] (PM1) -- (MMI5);
        
                    \node[block, right= 0.99cm of MMI5, yshift=-3.55cm, xshift=0.7cm] (MMI9) {$MMI_{2N+1}$};
        
                    \draw[waveguide] (MMI5.east)++(0,-0.25) -- ++(0.5,0) to[out=0, in=180] ++(1.2,-3);
        
                    \draw[waveguide] (MMI5.east) ++(0,0.25) -- ++(0.5,0) to[out=0, in=180] ++(1.2,3);
        
                    \node[block, below=of PD5, xshift=2.2cm, yshift=-7.15cm] (MMI6) {$MMI_{N+2}$};
                    
                    \draw[waveguide] (MMI6.east)++(0,-0.25) -- ++(0.5,0) to[out=0, in=180] ++(1.2,-3);
        
                    \draw[waveguide] (MMI6.east) ++(0,0.25) -- ++(0.5,0) to[out=0, in=180] ++(1.2,3);
        
                    \node[block, left=of MMI6, xshift=0.25cm] (PM2) {$PM_{2}$};
        
                    \draw[waveguide] (PM2) -- (MMI6);
        
                    \node[block, left=of PM2, yshift=2.4cm, xshift=-0.25cm] (MMI2) {$MMI_{2}$};
        
                    \draw[waveguide] (MMI2.east) ++(0,0.25) -- ++(0.5,0) to[out=0, in=180] ++(0.75,2);
                    
                    \draw[waveguide] (MMI2.east)++(0,-0.25) -- ++(0.5,0) to[out=0, in=180] ++(0.75,-2.125);
        
                    \node[photodiode, above=of MMI2, yshift=-2.9cm, xshift=2.45cm] (PD3) {$PD_{N+2}$};
        
                    \node[amplifier, shape border rotate=-90, scale=0.6, dashed, left=of MMI2, font=\fontsize{12}{10}\selectfont] (SOA2) {$SOA_{2}$};
                    
                    \draw[waveguide] (SOA2) -- (MMI2);
        
                    \node[left=of SOA2, font=\fontsize{10}{10}\selectfont\bfseries] (input2) {$B(\mathbf{r},t)$};
        
                    \draw[waveguide] (input2) -- (SOA2);
        
                    \node[photodiode2, right=of MMI9] (PD1) {$PD_{1}$};
        
                    \draw[waveguide] (MMI9) -- (PD1);
        
                    \node[block, right= 0.99cm of MMI6, yshift=-3.55cm, xshift=0.7cm] (MMI10) {$MMI_{2N+2}$};
        
                    \node[block, below=of PD3, xshift=1.9cm, yshift=-6.55cm] (MMI7) {$MMI_{N+3}$};
        
                    \draw[waveguide] (MMI7.east)++(0,-0.25) -- ++(0.5,0) to[out=0, in=180] ++(1.2,-3);
        
                    \draw[waveguide] (MMI7.east) ++(0,0.25) -- ++(0.5,0) to[out=0, in=180] ++(1.2,3);
        
                    \node[block, left=of MMI7, xshift=0.25cm] (PM3) {$PM_{3}$};
        
                    \draw[waveguide] (MMI7) -- (PM3);
        
                    \node[block, left=of PM3, yshift=2.4cm, xshift=-0.25cm] (MMI3) {$MMI_{3}$};
        
                    \draw[waveguide] (MMI3.east) ++(0,0.25) -- ++(0.5,0) to[out=0, in=180] ++(0.75,2);
                    
                    \draw[waveguide] (MMI3.east)++(0,-0.25) -- ++(0.5,0) to[out=0, in=180] ++(0.75,-2.125);
        
                    \node[photodiode, above=of MMI3, yshift=-2.9cm, xshift=2.45cm] (PD5) {$PD_{N+3}$};
        
                    \node[amplifier, shape border rotate=-90, scale=0.6, dashed, left=of MMI3, font=\fontsize{12}{10}\selectfont] (SOA3) {$SOA_{3}$};
        
                    \draw[waveguide] (SOA3) -- (MMI3);
        
                    \node[photodiode2, right=of MMI10] (PD2) {$PD_{2}$};
        
                    \draw[waveguide] (PD2) -- (MMI10);
        
                    \node[left=of SOA3, font=\fontsize{10}{10}\selectfont\bfseries] (input3) {$C(\mathbf{r},t)$};
        
                    \draw[waveguide] (input3) -- (SOA3);

                    \node[block2, below=of MMI7, yshift=-2.5cm] (MMI8) {$MMI_{2N}$};
        
                    \draw[waveguide] (MMI8.east)++(0,-0.25) -- ++(0.5,0) to[out=0, in=180] ++(1.2,-3);
        
                    \draw[waveguide] (MMI8.east) ++(0,0.25) -- ++(0.5,0) to[out=0, in=180] ++(1.2,3);

                    \node[block, left=of MMI8, xshift=0.25cm] (PM4) {$PM_{N}$};
        
                    \draw[waveguide] (PM4) -- (MMI8);
        
                    \node[block, left=of PM4, yshift=2.4cm, xshift=-0.25cm] (MMI4) {$MMI_{N}$};
        
                    \draw[waveguide] (MMI4.east) ++(0,0.25) -- ++(0.5,0) to[out=0, in=180] ++(0.75,2);
                    
                    \draw[waveguide] (MMI4.east)++(0,-0.25) -- ++(0.5,0) to[out=0, in=180] ++(0.75,-2.125);
        
                    \node[photodiode2, above=of MMI4, yshift=-2.9cm, xshift=2.45cm] (PD6) {$PD_{2N}$};
        
                    \node[amplifier, shape border rotate=-90, scale=0.6, dashed, left=of MMI4, font=\fontsize{12}{10}\selectfont] (SOA4) {$SOA_{N}$};

                    \draw[waveguide] (SOA4) -- (MMI4);
        
                    \draw[->] ($(SOA4)!0.55!(MMI4)$) ++(0,-1.5) -- ($(SOA4)!0.55!(MMI4)$)
                    node[pos=0, below] {waveguide};
        
                    \node[left=of SOA4, font=\fontsize{10}{10}\selectfont\bfseries] (input4) {$D(\mathbf{r},t)$};
        
                    \draw[waveguide] (SOA4) -- (input4);
        
                    \node[draw=none] (ellipsis1) at (4.9,-0.6-0.4) {$\vdots$};
        
                    \node[draw=none] (ellipsis1) at (4.9,-29.5-0.4) {$\vdots$};
        
                    \node[draw=none] (ellipsis1) at (4.9,-21.95-0.5) {$\vdots$};
        
                    \draw[waveguide, postaction={decorate, decoration={markings, mark=at position 0.5 with {\arrow{<}}}}, line width=0.25pt] (SOA1) -- ++(0,1) node[above, font=\fontsize{10}{10}\selectfont] {$I_{dc1}$};
        
                    \draw[waveguide, postaction={decorate, decoration={markings, mark=at position 0.5 with {\arrow{<}}}}, line width=0.25pt] (SOA2) -- ++(0,1) node[above, font=\fontsize{10}{10}\selectfont] {$I_{dc2}$};
        
                    \draw[waveguide, postaction={decorate, decoration={markings, mark=at position 0.5 with {\arrow{<}}}}, line width=0.25pt] (SOA3) -- ++(0,1) node[above, font=\fontsize{10}{10}\selectfont] {$I_{dc3}$};
        
                    \draw[waveguide, postaction={decorate, decoration={markings, mark=at position 0.5 with {\arrow{<}}}}, line width=0.25pt] (SOA4) -- ++(0,1) node[above, font=\fontsize{10}{10}\selectfont] {$I_{dc4}$};
        
                    \draw[waveguide, postaction={decorate, decoration={markings, mark=at position 0.5 with {\arrow{>}}}}, line width=0.25pt] (PD4) -- ++(0,1) node[above, font=\fontsize{7}{10}\selectfont] {Photometry N+1};
        
                    \draw[waveguide, postaction={decorate, decoration={markings, mark=at position 0.5 with {\arrow{>}}}}, line width=0.25pt] (PD3) -- ++(0,1) node[above, font=\fontsize{7}{10}\selectfont] {Photometry N+2};
        
                    \draw[waveguide, postaction={decorate, decoration={markings, mark=at position 0.5 with {\arrow{>}}}}, line width=0.25pt] (PD5) -- ++(0,1) node[above, font=\fontsize{7}{10}\selectfont] {Photometry N+3};
        
                    \draw[waveguide, postaction={decorate, decoration={markings, mark=at position 0.5 with {\arrow{>}}}}, line width=0.25pt] (PD6) -- ++(0,1) node[above, font=\fontsize{7}{10}\selectfont] {Photometry 2N};
        
                    \draw[waveguide, postaction={decorate, decoration={markings, mark=at position 0.5 with {\arrow{>}}}}, line width=0.25pt] (PD1) -- ++(0,1) node[above, font=\fontsize{7}{10}\selectfont] {Photometry 1};
        
                    \draw[waveguide, postaction={decorate, decoration={markings, mark=at position 0.5 with {\arrow{>}}}}, line width=0.25pt] (PD2) -- ++(0,1) node[above, font=\fontsize{7}{10}\selectfont] {Photometry 2};
        
                    \draw[waveguide, postaction={decorate, decoration={markings, mark=at position 0.5 with {\arrow{<}}}}, line width=0.25pt] (PM1) -- ++(0,1) node[above, font=\fontsize{10}{10}\selectfont] {$I_{dc_{N+1}}$};
        
                    \draw[waveguide, postaction={decorate, decoration={markings, mark=at position 0.5 with {\arrow{<}}}}, line width=0.25pt] (PM2) -- ++(0,1) node[above, font=\fontsize{10}{10}\selectfont] {$I_{dc_{N+2}}$};
        
                    \draw[waveguide, postaction={decorate, decoration={markings, mark=at position 0.5 with {\arrow{<}}}}, line width=0.25pt] (PM3) -- ++(0,1) node[above, font=\fontsize{10}{10}\selectfont] {$I_{dc_{N+3}}$};
        
                    \draw[waveguide, postaction={decorate, decoration={markings, mark=at position 0.5 with {\arrow{<}}}}, line width=0.25pt] (PM4) -- ++(0,1) node[above, font=\fontsize{10}{10}\selectfont] {$I_{dc_{2N}}$};
        
                    \draw [waveguide, line width=0.25pt] (PD1.south) -- ++(0,-0.25) node[ground, below, rotate=270, scale=0.7, xshift=0.1cm, yshift = 0.25cm] {};
        
                    \draw [waveguide, line width=0.25pt] (PD2.south) -- ++(0,-0.25) node[ground, below, rotate=270, scale=0.7, xshift=0.1cm, yshift = 0.25cm] {};
        
                    \draw [waveguide, line width=0.25pt] (PD3.south) -- ++(0,-0.25) node[ground, below, rotate=270, scale=0.7, xshift=0.1cm, yshift = 0.25cm] {};
        
                    \draw [waveguide, line width=0.25pt] (PD4.south) -- ++(0,-0.25) node[ground, below, rotate=270, scale=0.7, xshift=0.1cm, yshift = 0.25cm] {};
        
                    \draw [waveguide, line width=0.25pt] (PD5.south) -- ++(0,-0.25) node[ground, below, rotate=270, scale=0.7, xshift=0.1cm, yshift = 0.25cm] {};
        
                    \draw [waveguide, line width=0.25pt] (PD6.south) -- ++(0,-0.25) node[ground, below, rotate=270, scale=0.7, xshift=0.1cm, yshift = 0.25cm] {};
        
                    \draw [waveguide, line width=0.25pt] (SOA1.south) -- ++(0,-0.25) node[ground, below, rotate=270, scale=0.7, xshift=0.1cm, yshift = 0.25cm] {};
        
                    \draw [waveguide, line width=0.25pt] (SOA2.south) -- ++(0,-0.25) node[ground, below, rotate=270, scale=0.7, xshift=0.1cm, yshift = 0.25cm] {};
        
                    \draw [waveguide, line width=0.25pt] (SOA3.south) -- ++(0,-0.25) node[ground, below, rotate=270, scale=0.7, xshift=0.1cm, yshift = 0.25cm] {};
        
                    \draw [waveguide, line width=0.25pt] (SOA4.south) -- ++(0,-0.25) node[ground, below, rotate=270, scale=0.7, xshift=0.1cm, yshift = 0.25cm] {};
        
                    \draw [waveguide, line width=0.25pt] (PM1.south) -- ++(0,-0.25) node[ground, below, rotate=270, scale=0.7, xshift=0.1cm, yshift = 0.25cm] {};
        
                    \draw [waveguide, line width=0.25pt] (PM2.south) -- ++(0,-0.25) node[ground, below, rotate=270, scale=0.7, xshift=0.1cm, yshift = 0.25cm] {};
        
                    \draw [waveguide, line width=0.25pt] (PM3.south) -- ++(0,-0.25) node[ground, below, rotate=270, scale=0.7, xshift=0.1cm, yshift = 0.25cm] {};
        
                    \draw [waveguide, line width=0.25pt] (PM4.south) -- ++(0,-0.25) node[ground, below, rotate=270, scale=0.7, xshift=0.1cm, yshift = 0.25cm] {};

                \end{tikzpicture}
            }
        \end{minipage}
    }

    \caption{Proposed IP-WFS circuit. \textit{Top}: Simplified circuit diagram illustrating how the wavefront sensing is performed through the phenomenon of interferometry, resulting in an electrical signal. \textit{Bottom}: Complete circuit diagram, including all the components.}
    \label{fig:ipwfs_circuits}
\end{figure}

A system-level implementation of this configuration is presented in the upper panel of Fig. \ref{fig:ipwfs_circuits}. The block $\varphi$ is responsible for performing the interferometry between the different inputs of the system, resulting in the photometric output of each interferometry measured by a photodiode (PD). The lower panel shows a detailed block diagram of the proposed method to measure phase shifts. In this circuit, semiconductor optical amplifiers (SOAs) can be optionally added to each input. These amplifiers dynamically adjust their gain to ensure that the two inputs have the same amplitude, which makes data analysis easier. Additionally, these amplifiers can provide a higher dynamic range, enhancing the sensitivity of the sensor \citep{6069519}. The main issue of SOAs are their non-linearities and high power consumption. Then, by using multimode interferometers (MMIs) that act as beam splitters, the light is split in a certain ratio. Part of the light is sent to a photodetector to perform photometry, sensing the light intensity at the input. The remainder is sent to another MMI. This  allows, for example,  input $B(\mathbf{r},t)$ to be compared with inputs $A(\mathbf{r},t)$ and $C(\mathbf{r},t)$.

As previously explained, SOAs can offer several advantages, such as simplifying data analysis or providing a higher dynamic range. However, removing them can also be advantageous, for example by simplifying the design, reducing power and area consumption, and reducing the number of control signals. Without SOAs, photometry information provides the intensity of the input signals. In combination with the output signals, the phase differences can be estimated.

One problem that arises with this configuration is that interferometry is a technique that relies on the symmetry of the optical paths followed by the signals that are going to be compared. If one path is slightly longer than the other during the fabrication process, this introduces a systematic error in the collected data. To correct this, a phase modulator (PM) is introduced for each input to perform a calibration. During calibration, a set of already known optical signals with the same amplitude and phase is introduced to each input. The phase modulators then modify the phase of each input until a maximum is reached at each of the photodetectors, thus correcting the delays introduced during the fabrication process.

Table \ref{Components} provides a summary of the number of components required as a function of the number of inputs, N. As can be seen, removing the SOAs results in a reduction of N components and 2N electrical connections, reducing the area and power consumption.

\begin{table}
\caption{List of components.}  
\label{Components}
\centering                         

\begin{tabular}{c c c c c}        

\hline\hline                 
 & PMs & MMIs & SOAs & PDs \\    
\hline                        
   Components & N & 3N + 1 & N & 2N + 1\\
   El. connection & 2N & 0 & 2N & 2(2N + 1) \\
\hline                                   

\end{tabular}
\tablefoot{PMs: phase modulators, MMIs: multimode interferometers, SOAs: semiconductor optical amplifiers, PDs: photodetectors.}
\end{table}

\subsection{Mathematical description}

This WFS is analysed mathematically based on the method presented in the book \citep{capmany2020programmable}, which describes each of the different elements commonly used in integrated photonics by the scattering matrix $\mathbf{S}$, as $\mathbf{b} = \mathbf{S}\mathbf{a}$, where $\mathbf{a}$ and $\mathbf{b}$ are column vectors representing the inputs and outputs, respectively. 

A waveguide is generally employed to transport light within the circuit, enabling light to be confined within its boundaries through the phenomenon of total internal reflection \citep{doi:https://doi.org/10.1002/0470861401.ch3}. In Fig. \ref{fig:ipwfs_circuits} it is represented by the connections between the components. The scattering matrix for a single-mode waveguide can be simply defined as $\mathbf{S_{WG}}=\mathbbm{1}$.

As previously mentioned, SOAs can be optionally added to enhance the dynamic range by maintaining the same amplitude level for both inputs. The scattering matrix of an SOA can be defined as $\mathbf{S_{SOA}}=\sqrt{G_{\rm SOA} }$, where $G_{\rm SOA}$ refers to the gain provided by the SOA. For the rest of the analysis, $G_{\rm SOA}$ is considered as 1.

The subsequent element within the circuit is a multimode interferometer (MMI). Only 1x2 and 2x1 MMI configurations were used for this specific work, so the analysis was performed for these particular arrangements. This component splits the light in an R:1-R ratio, where R ranges from 0 to 1. Some of the light is directed to a photodiode, which generates a current proportional to the light intensity. This current signal is a direct measurement of the light intensity at the input of the wavefront sensor, and can be used to control the gain of the SOAs. 

The resulting S-matrix of an MMI is presented in Eq. \eqref{Smatrix_repr}, where internal reflections have been neglected; an analysis of these reflections was conducted in \cite{300172}:

{\begin{equation}
    \mathbf{S_{MMI}} =
    \begin{pmatrix}
        0 & \frac{1}{\sqrt{2}} & \frac{1}{\sqrt{2}} \\
        \sqrt{R} & 0 & 0 \\
        \sqrt{1-R} & 0 & 0  \\
    \end{pmatrix}.
    \label{Smatrix_repr}
\end{equation}}

As in our case MMIs are used either as 50:50 beamsplitter (1x2 MMI) or as interferometers (2x1 MMI), the matrix \eqref{Smatrix_repr} can split into matrices \eqref{Smatrix_1x2} and \eqref{Smatrix_2x1}, respectively: 

{\begin{equation}
    \mathbf{S_{MMI_{1x2}}} =
    \begin{pmatrix}
        \sqrt{R} \\
        \sqrt{1-R} \\
    \end{pmatrix},
    \label{Smatrix_1x2}
\end{equation}}

{\begin{equation}
     \mathbf{S_{MMI_{2x1}}} = 
    \begin{pmatrix}
        \frac{1}{\sqrt{2}} & \frac{1}{\sqrt{2}} \\
        0 & 0 \\
        0 & 0 \\
    \end{pmatrix}.
    \label{Smatrix_2x1}
\end{equation}}

To perform a perfect interferometry, it is essential that the paths followed by all the inputs are identical. In practice, the manufacturing process makes it challenging to achieve, due to process tolerances. In order to achieve near perfect interferometry, some phase modulators (PMs) are added. Any discrepancies in the paths will result in phase differences between the signals; thus, a calibration stage should be carried out before starting to work with the wavefront sensor. A general S-matrix for performing this phase shift is described in Eq. \eqref{eq:Phase-Shifter},

{\begin{equation}
\mathbf{S_{PM}} = \begin{pmatrix}
    0 & e^{-j\theta(I_{\rm dc})} \\
    e^{-j\theta(I_{\rm dc})} & 0
\end{pmatrix},
\label{eq:Phase-Shifter}
\end{equation}}

\noindent where the factor $\theta(I_{dc})$ represents the phase shift introduced by the PM, which depends on the polarisation current used.

There are two principal methods for physically implementing this element. One of the most popular implementations in integrated photonics is based on the electro-optic or Pockels effect \citep{9286547,paschotta2008encyclopedia}. This type of modulator has the advantage of offering a large modulation bandwidth \citep{Xu:22}; however, it also exhibits a notable insertion loss of up to 6 dB \citep{Qiao2017}. Another type of phase modulator is the thermo-optic phase modulator \citep{Liu2022}, which exhibits insertion losses of typically 1-2 dB/cm. Since the sampling frequencies in solar adaptive optics are of the order of kHz \citep{Rao2024}, there is no need to integrate modulators with high modulation bandwidth. Given the current state of integrated photonic technologies, thermo-optic phase modulators are better suited for this particular application, due to their low insertion losses.

Finally, the optical signal will arrive at the photodetector, which will convert it into an electrical signal. Since photodetectors are usually not capable of detecting the frequency of light, they only observe the intensity ($I(\mathbf{r})$) on an optical signal \citep{paschotta2008encyclopedia}.

This mathematical description presents an updated model for an arbitrary number of pixels with respect to the work presented in \citep{10966318}. Equation \eqref{eq:signal_description} represents the intensity that the photodiode will receive after performing the interferometry,

\begin{equation}
I_{P}(\mathbf{r}) = 2I(\mathbf{r}) + 2I(\mathbf{r})\cos(\Delta \phi)
\label{eq:signal_description}
,\end{equation}

\noindent where $\Delta \phi = \phi_{1} - \phi_{2}$ corresponds to Eq. \eqref{varphi}, and $I_{P}(\mathbf{r})$ is the optical intensity that arrives at the photodetector.  This intensity will depend upon the phase shift, denoted by $\cos(\phi_1 - \phi_2)$, and shows a non-linear behaviour of the WFS:

\begin{equation}
\begin{split}
\Delta \phi 
&= (\sigma_{1} - \sigma_{2}) + (\theta_{1}(I_{\rm dc2}) - \theta_{2}(I_{\rm dc4}))
\end{split}
\label{varphi}
.\end{equation}

\noindent This equation represents the phase shift that will be detected by the WFS, where $\sigma_{1}$ and $\sigma_{2}$ represent the original phases at each of the inputs. Ideally, the phase shift presented in the wavefront at the input of the WFS, $\sigma_{1} - \sigma_{2}$, would be equal to the actual phase shift. However, due to the aforementioned manufacturing process mismatches, this value may differ. To solve it, during the calibration stage, the phase modulators should introduce a phase shift $\theta_{1}(I_{\rm dc2}) - \theta_{2}(I_{\rm dc4})$ that compensates for any path mismatches that could be introduced during the fabrication process.

This light intensity can be converted into optical power using the relationship $P(\mathbf{r})=I(\mathbf{r})A_{\rm eff}$, where $A_{\rm eff}$ is the effective area. As the input effective areas and the output are assumed to remain equal, in this case, by using a waveguide of the same size, the optical power should be equal to the intensity. The electrical intensity that the photodetector will deliver  depends upon the quantum efficiency (QE). This magnitude describes the efficiency of the conversion of incident photons into electrons.

Up to this point, this wavefront sensor is capable of measuring the absolute phase differences between two points of the wavefront. However, it does not directly measure which of the two points in the wavefront is delayed with respect to the other since the cosine function is symmetric. In order to measure this, a two-measurement strategy is adopted. First, the WFS measures atmospheric aberrations normally, thus obtaining the absolute value of the phase difference. Afterward, a small phase shift is introduced in one of the two inputs, for example $\phi_{1}$, and another measurement is recorded. 

\begin{figure}[h]
   \centering
   \resizebox{0.44\textwidth}{!}{
      \includegraphics{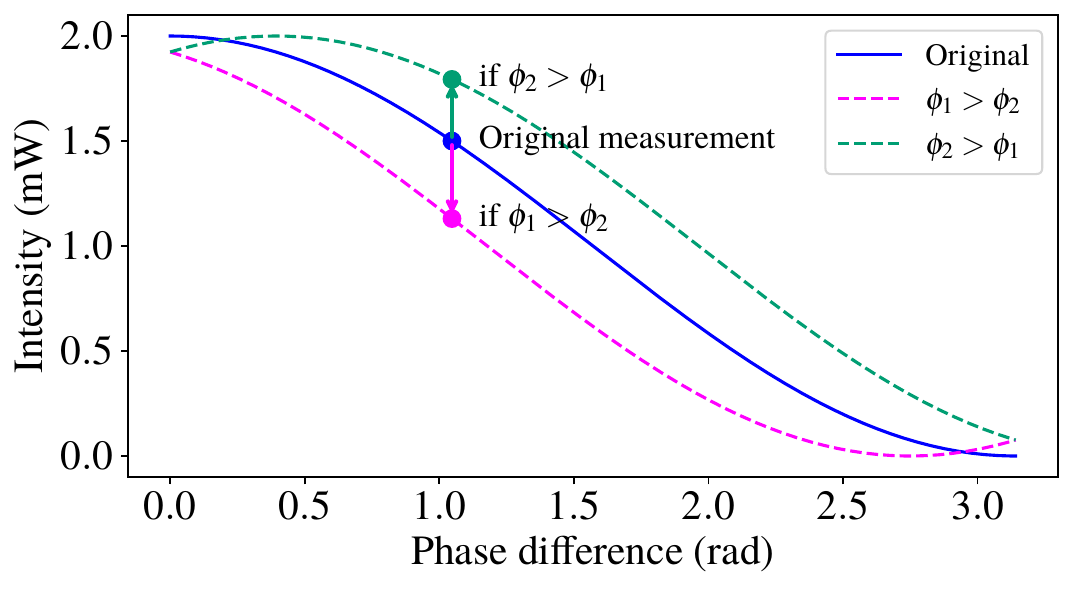}}
   \caption{Original and after calibration measurements. Firstly, a direct measurement is done, which yields the absolute value of the phase difference. To know whether $\phi_{1}$ is bigger or smaller than $\phi_{2}$, a second measurement is taken, thus adding a phase shift to $\phi_{1}$.}
   \label{Calibration_explaination}
\end{figure}

Once this data is recorded, two scenarios can be present. The first scenario is that the interferometry measured after the second value is recorded is bigger than the original one. This can only occur if $\phi_{1} > \phi_{2}$, and vice versa, if the recorded value is lower, it means that $\phi_{2} > \phi_{1}$. It is  important to note  that to be able to make these measurements, the data recording should be faster than the changes experimented in the wavefront. An example of this effect is illustrated in Fig. \ref{Calibration_explaination}. 

There is an alternative method of performing wavefront sensing using only one measurement, which will double the frequency at which the adaptive optics system could work. It could be achieved by inserting an initial phase shift of $\pi/2$ to $\theta_{2}(I_{\rm dc4})$ or $\theta_{4}(I_{\rm dc4})$. Then Eq. \eqref{eq:signal_description} will involve a sine instead of a cosine. In this case, the wavefront sensor will work with only one measurement in the range $-\pi/2 < \sigma_{1} - \sigma_{2} < \pi/2$. Outside of that range, there will always be at least two possible values of phase difference for each intensity value.

\section{Integrated photonic chip development platforms}

\label{Integrated photonic chip development platforms}

One of the key aspects of the design of a photonic integrated circuit is the selection of an appropriate material for the intended design. Silicon is of particular interest, due our ability to leverage the extensive knowledge acquired in the microelectronics industry, thereby allowing robust and low-cost processing. It can also easily be integrated with complementary metal–oxide–semiconductor (CMOS) technology. \citep{Thomson_2016, Siew:21}. 

Furthermore, silicon exhibits favourable optical properties, including low losses across a range of wavelengths and a compact form factor, which enables the miniaturisation of photonic elements and increases the density of components on a single chip \citep{10216293}. However, the large-scale integration of active components is still a challenge. Silicon photonics has seen considerable advancements, with silicon-on-insulator (SOI) emerging as the most extended and widely used approach, due to its large-scale integration, the large number of foundries available, low propagation losses, and the number of chips that can be placed per wafer. 

In recent years, silicon nitride (Si$_3$N$_4$) has become an emerging candidate due to its large transparency window. This platform offers waveguides with almost no propagation losses and robustness to manufacturing variations, with the drawback of having more voluminous waveguides \citep{rinaldi:tel-03639080}.

In addition to silicon photonics, indium phosphide (InP) represents another prominent platform, given its capacity to combine passive and active components on a single chip, without the need for hybrid integration \citep{rinaldi:tel-03639080}. The direct use of InP as a platform enhances the performance of the active components compared to hybrid integration. However, this approach has been found to decrease the performance of the passive components. Several designs have already been developed using this technology \citep{10041912}, some of them in the field of astronomy \citep{10261_349694}. Table \ref{Performance-summary} provides a summary of the properties of the different materials presented. This table is based on data collected from \citep{rinaldi:tel-03639080} and the Jeepix website.\footnote{\href{https://www.jeppix.eu/workflow/performance-summary-table/}{https://www.jeppix.eu/workflow/performance-summary-table/}}

\begin{table}
\caption{Comparison of different IP platforms.}  
\label{Performance-summary}
\label{table:1}      
\centering                        
\small
\begin{tabular}{c c c c}        
\hline\hline                 
 & SOI & Si$_3$N$_4$ & InP \\    
\hline                        
   PM & \checkmark & Modest & \checkmark\\
   MMI & \checkmark & \checkmark & \checkmark \\
   SOA & Hybrid/Heter. & Hybrid/Heter. & \checkmark \\      
   Laser & Hybrid/Heter. & Hybrid/Heter. &  \checkmark \\
   Photodiodes & Hybrid/Heter. & Hybrid/Heter. &  \checkmark \\
   Light coupling & EC/SSC/Gr. & EC/SSC/Gr. & EC/SSC \\
   Foundries & Several & Few & Several \\
   Propagation loss & Low & Almost any & Low/Med. \\ 
\hline                                   

\end{tabular}
\tablefoot{EC: edge coupling, SSC: spot-size converter, Gr.: grating couplers.}
\end{table}

\section{Light coupling}

\label{Light coupling}

One of the most critical and intricate challenges in astrophotonics is the efficient coupling of light between different optical stages. In the case of solar adaptive optics, coupling efficiency is of particular importance, since a large number of corrective elements are required due to the worse seeing conditions during the day time and the fact that most of the science is done at visible wavelengths \citep{Rimmele2011}, which requires a high density of pixels in the wavefront sensor. This means that the light present in the wavefront must be divided by the total number of pixels, which, in conjunction with the originally extremely low coupling efficiency and the optical losses that are produced inside the circuit, could result in a low photon flux that the photodetectors can use to perform wavefront sensing.

For infrared observations, the conditions are even worse since the amount of photons is considerably lower with respect to the visible light for solar observations, and there are also fewer absorption lines, which are normally weaker \citep{Penn2014}. This results in a reduction in the signal-to-noise ratio of the measured signals. As the presented WFS is based on interferometry, this effect increases the wavefront error (WFE).

This section explores the coupling mechanisms between telescopes, optical fibres, and photonic chips. We focus on two key processes: telescope to optical fibre coupling and optical fibre to photonic chip coupling.

\subsection{Optical fibre to chip coupling}

\begin{figure}[h]
    \centering
    \includegraphics[width=0.25\textwidth]{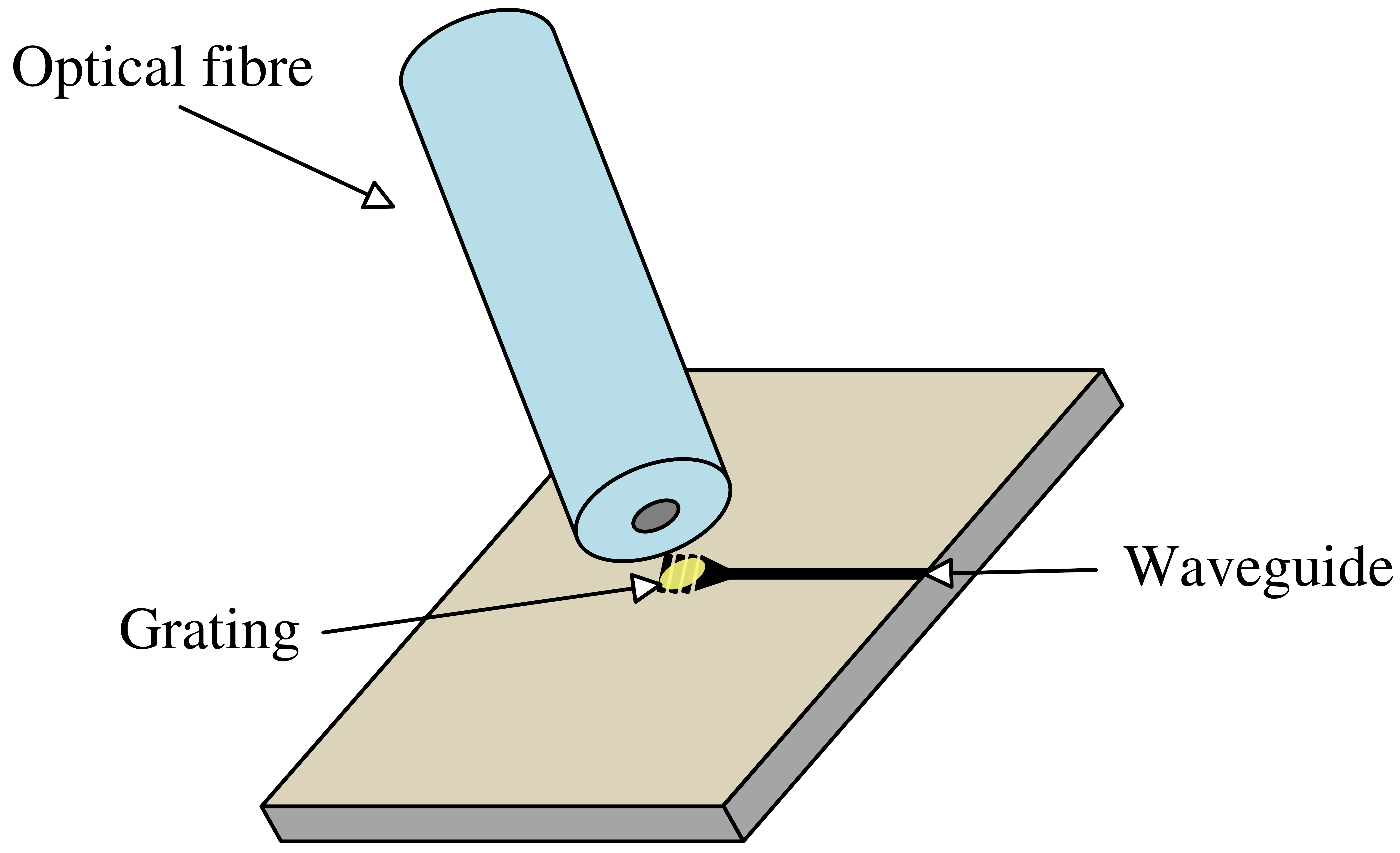}
    \hfill
    \includegraphics[width=0.23\textwidth]{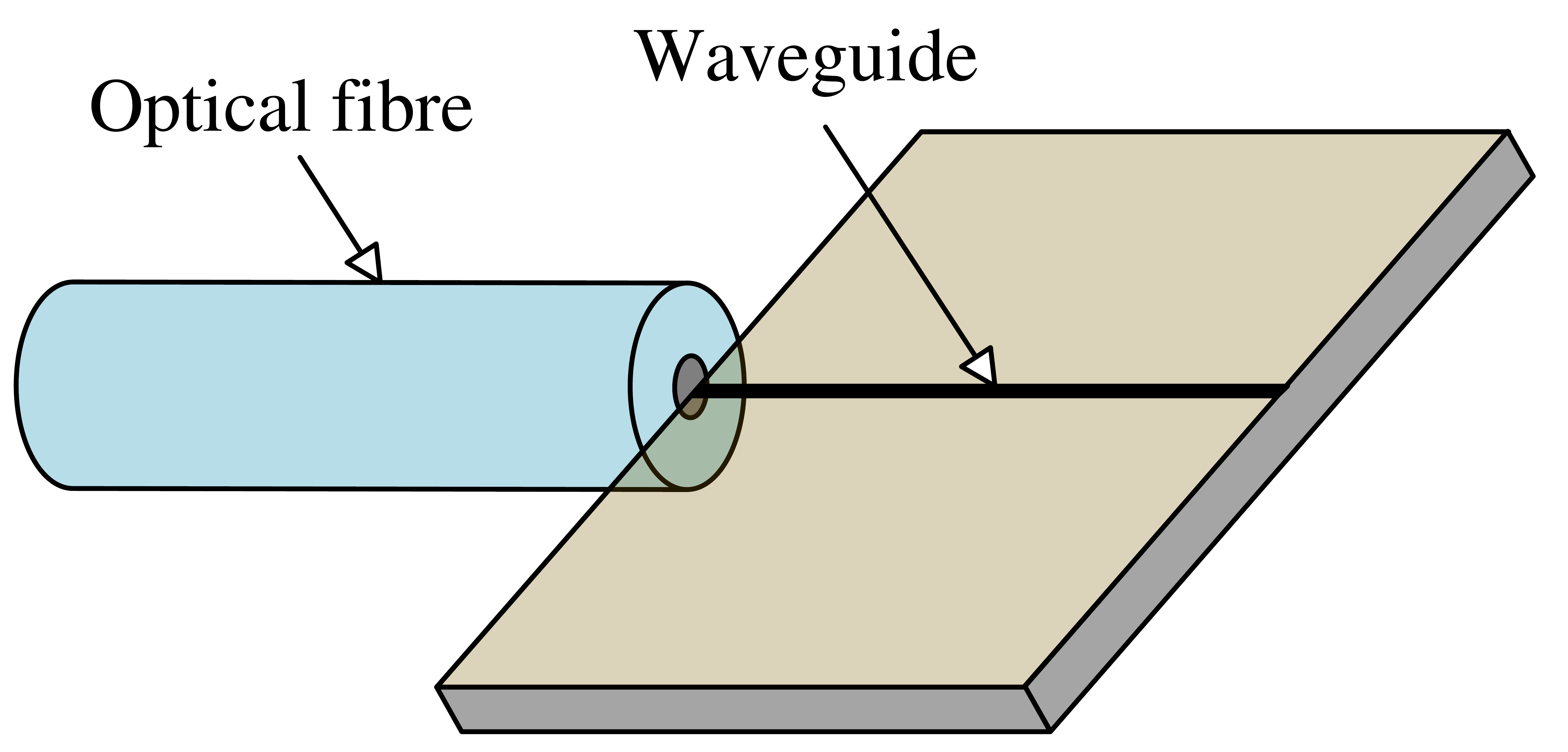}
    \caption{Two different methods for performing light coupling to PICs. \textit{Left}: Light coupling using grating couplers. \textit{Right}: Light coupling using edge couplers.}
    \label{Grating-couplers}
\end{figure}

In the field of astronomy, grating couplers represent a particularly intriguing method. This element is constituted of a periodic structure of different materials arranged on a given surface (see Fig. \ref{Grating-couplers}). An in-depth analysis of the underlying physics behind this device is provided in \citep{mi11070666}. An additional option for light coupling into the chip is the use of edge couplers, which consist of an in-plane coupling, where the optical fibre and a waveguide positioned at the edge of the PIC are aligned \citep{app10041538} (see Fig. \ref{Grating-couplers}). A particular case of edge couplers is the spot-size converter (SSC).

Edge couplers are a widely used option for coupling light as they can be integrated into platforms such as InP. However, grating couplers emerge as a potential and favourable option for this kind of application due to their low losses and the possibility of directly coupling the light using a microlens array without the need to use optical fibres.

\subsection{Telescope to optical fibre coupling}

The issue of light coupling from a telescope to single-mode optical fibres is a problem that has been widely discussed in the literature (see \citep{Jovanovic}). The coupling efficiency is described by the overlap integral, Eq. \eqref{Overlap},

\begin{equation}
    \mu = \frac{\left| \int E_{0} E^*_{1} \, dA \right|^2}{\int |E_{0}|^2 \, dA \int |E_1|^2 \, dA},
    \label{Overlap}
\end{equation}

\noindent where $E_{0}$ is the incoming electric field from the telescope, and $E^*_{1}$ is the complex conjugate of the electric field distribution ($E_{1}$) of the optical fibre. 

\begin{figure}[h]
   \centering
   \resizebox{0.44\textwidth}{!}{
      \includegraphics{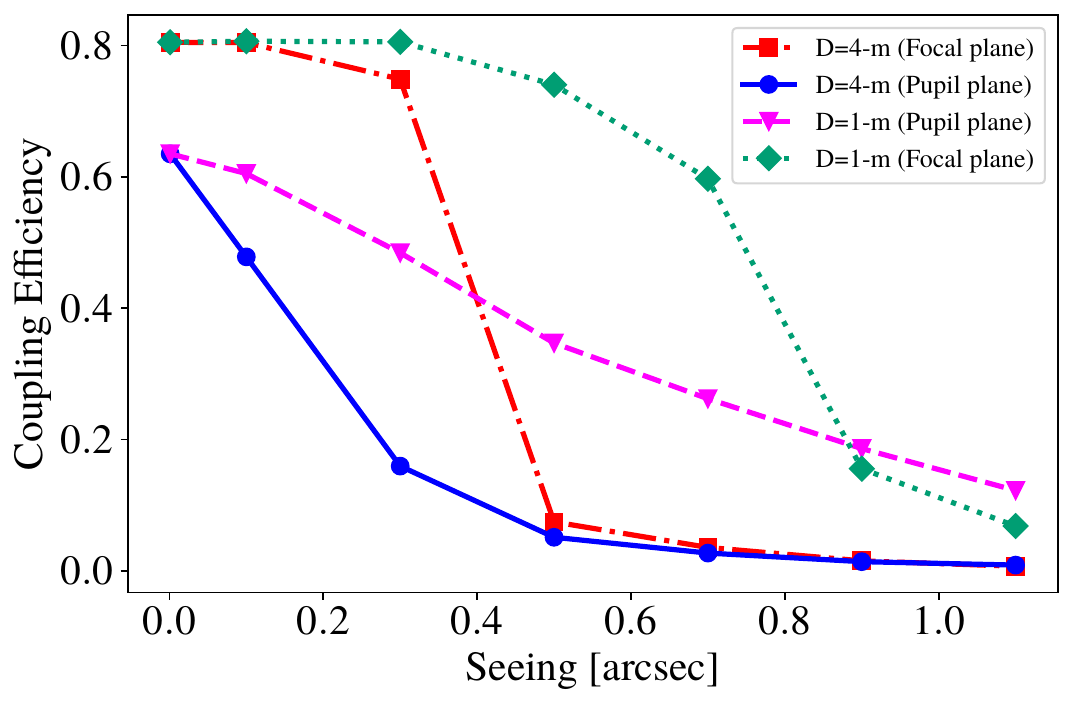}}
   \caption{Optical fibre coupling efficiency at the pupil and focal plane for a 1m and 4m class telescope for different seeing values at 1550 nm.}
   \label{Coupl_eff_pupil}
\end{figure}

The results presented in \citep{Ellis:21} show a series of calculations and simulations of the coupling efficiency for an 8m class telescope, in both the pupil and the focal plane. The results show that for realistic seeing conditions of more than 0.5 arcsec at a wavelength of 1550 nm, the coupling efficiency drops to almost $0$$\%$ in both cases. Based on these calculations, an analysis was performed to compare the results of a 4m class telescope with a 1m class telescope. The results are shown in Fig. \ref{Coupl_eff_pupil}. 

In the pupil plane, the simulation was carried out by applying a Kolmogorov phase screen ($\phi(k)$) to the telescope pupil in the form of $e^{j\phi(k)}$. The electric field distribution was then scaled to the appropriate pupil diameter and compared with the Gaussian modal profile of a single mode fibre using the overlap integral. This simulation was performed using 100 different phase screen distributions, for seeing values of 0.001, 0.1, 0.3, 0.5, 0.7, 0.9, and 1.1 arcsec with an outer scale ($L_{0}$) of 15 metres, and averaging. 

In the focal plane, the pupil electric field distribution was Fourier transformed to obtain its point-spread function (PSF). This simulation was performed for 100 realisations of the pupil electric field distribution. These simulations assume a perfect match between the f-number of the telescope and the numerical aperture (NA) of the optical fibre. The Kolmogorov phase screen was generated using the Python-based software AOTools \citep{Townson:19}. The results are shown in Fig. \ref{Coupl_eff_pupil} for 4m and 1m class telescopes.

The simulation shows an enhancement in the coupling efficiency for small telescopes, with a coupling efficiency of $\approx 40$$\%$ for seeing conditions < 0.5 arcsec in the pupil plane. This is mainly because the Fried parameter is smaller than the size of the telescope for seeing conditions < 0.31 arcsec at a wavelength of 1550 nm, which is the common operating wavelength for commercial integrated photonics. Meanwhile, for large telescopes, such as 4m class telescopes, this limit is found at just 0.078 arcsec, causing the coupling efficiency to decay quite rapidly with seeing.

This coupling efficiency can be improved by using adaptive optics or photonic lanterns \citep{doi:10.1142/q0391}. In a closed-loop AO system, the incoming beam is progressively corrected by the deformable mirror, thereby increasing the coupling efficiency into the single-mode fibres. However, during the initial iterations of the control loop, the incoming wavefront remains either totally or partially uncorrected. This results in a low coupling efficiency when the entire pupil is coupled into an SMF. This section proposes that splitting the pupil into different subapertures improves the coupling efficiency, even with an uncorrected wavefront.

Figure \ref{Coupl_eff_pupil} shows that the coupling efficiency is highly dependent on the size of the telescope aperture. To improve this coupling efficiency, this work proposes using an array of microlenses to couple the light to an array of optical fibres, instead of a single one. The spot size diameter of an airy disc depends on Eq. \eqref{eq:spot-size},

\begin{equation}
    \diameter_{PSF} = 2.44 \cdot \lambda \cdot F
\label{eq:spot-size},
\end{equation}

\noindent where F is the f-number, which is the relationship between the focal ratio (f) and the telescope diameter (D). 

This spot size should be matched to the mode field diameter (MFD) of an optical fibre, i.e. the point at which the overlap integral reaches a maximum. The relationship between the Fried parameter ($r_{0}$) and the seeing in arcsec is shown in Eq. \eqref{eq:r_0} \citep{seeing}: 

\begin{equation}
    r_{0} = 2.013\cdot10^5 \cdot \frac{\lambda}{seeing}
\label{eq:r_0}.
\end{equation}

Therefore, for small values of D, the effects of atmospheric turbulence seem to be reduced, and higher coupling efficiencies can be achieved. On the other hand, smaller aperture diameters create larger point spread functions (PSF), so the focal ratio should be selected to match an F-number close to the fibre NA value. 

\begin{figure}[h]
   \centering
   \resizebox{0.44\textwidth}{!}{
      \includegraphics{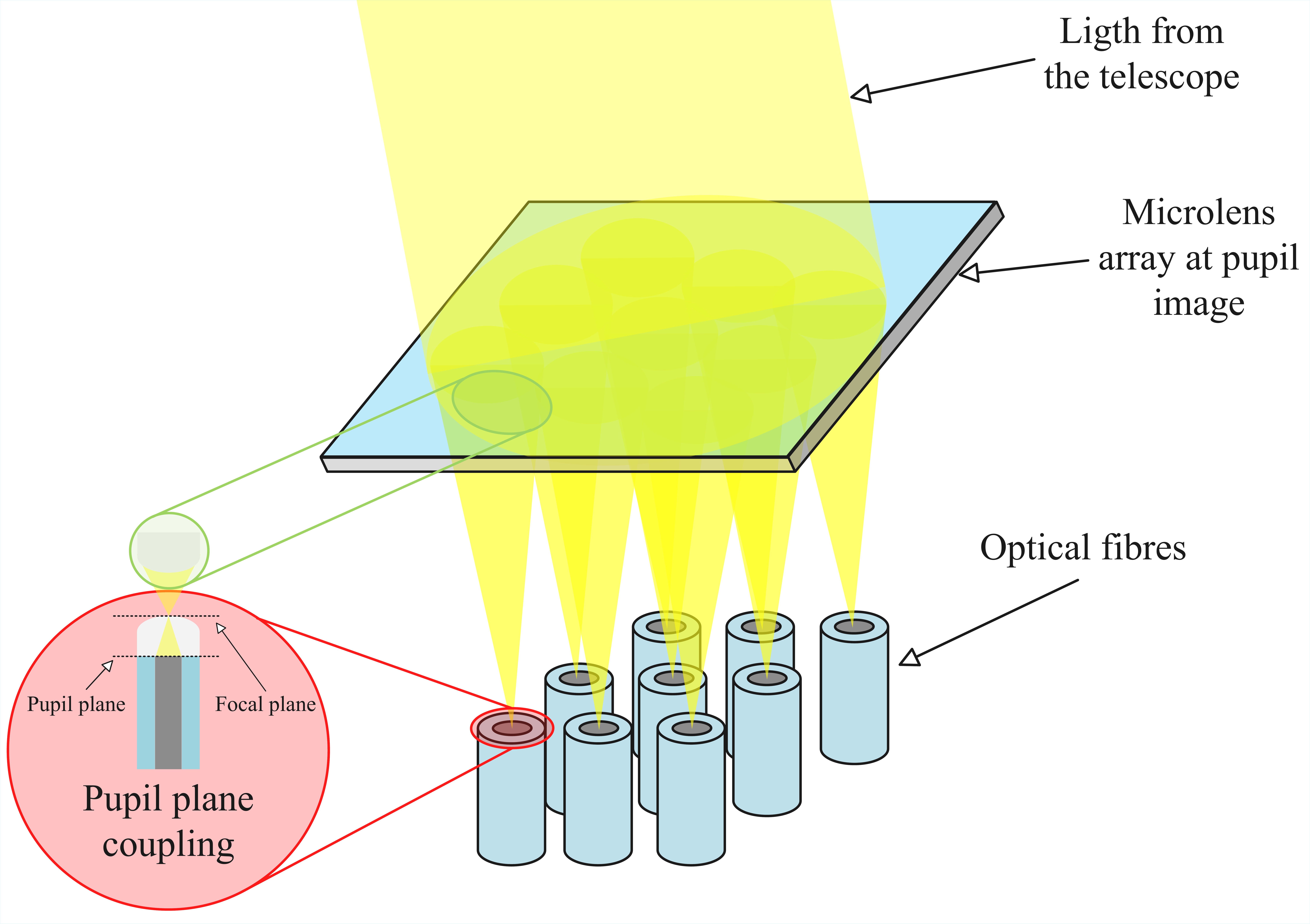}}
   \caption{Light coupling into an array of single-mode fibres using an array of microlenses. The pupil plane can be coupled by placing a microlense in the tip of the fibre \citep{Ellis:21}.}
   \label{PIC_Coupling}
\end{figure}

Consequently, the coupling efficiency is inversely proportional to the aperture size. This efficiency can be significantly improved by placing an array of microlenses in the pupil plane to couple a portion of the wavefront to an optical fibre. Another microlens placed at the tip of an optical fibre can couple a slice of the wavefront to an individual fibre. Therefore, to cover the entire wavefront, an array of optical fibres is required (see Fig. \ref{PIC_Coupling}).

\begin{figure}[h]
    \centering
    \includegraphics[width=0.44\textwidth]{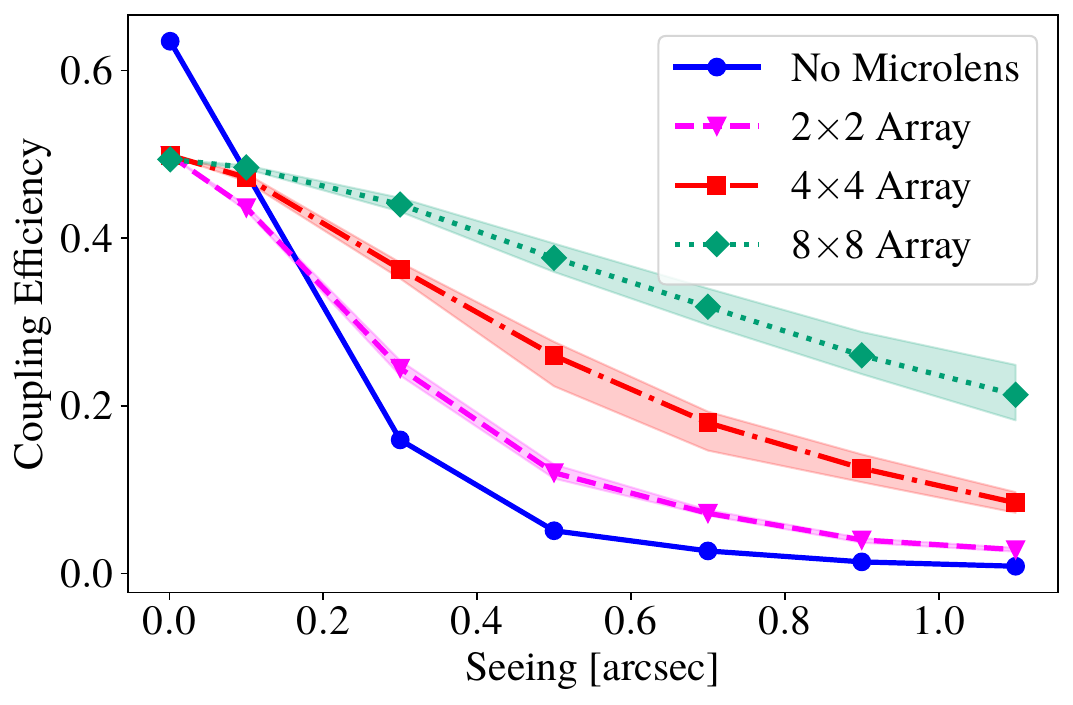}
    \hfill
    \includegraphics[width=0.44\textwidth]{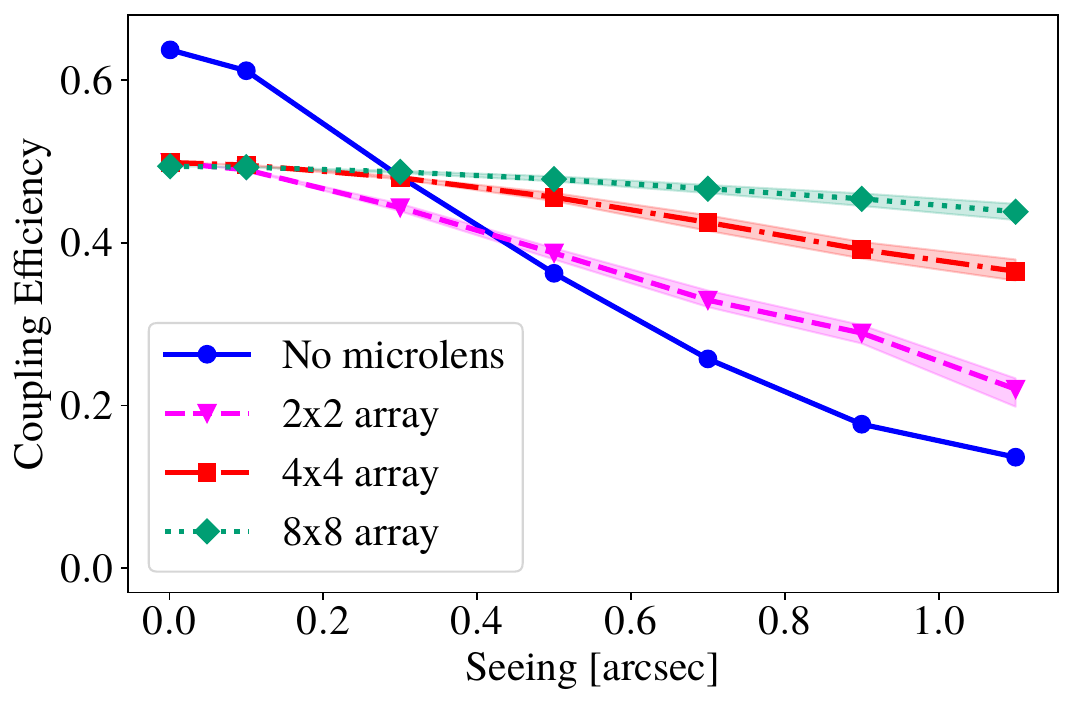}
    \caption{Total coupling efficiency for a 2×2, 4×4, and 8×8 microlens array, and without microlens array, in the pupil plane for a 4m class telescope (\textit{top}) and a 1m class telescope (\textit{bottom}).}
    \label{fig:total_coupling_efficiency}
\end{figure}

Figure \ref{fig:total_coupling_efficiency} depicts the total coupling efficiency for a 1m and 4m class telescope, respectively, both without a microlens array and with 2x2, 4x4, and 8x8 microlens arrays in the pupil plane. The fill space represents the averaged maximum and minimum coupling efficiency that each of the individual optical fibres can achieve.

It shows how the coupling efficiency increases for seeing values greater than 0.1 arcsec for 4m class telescopes and 0.4 arcsec for 1m class telescopes when an array of microlenses is introduced. For 4m class telescopes, it shows a coupling efficiency of over $35\%$ for an 8x8 microlens array at a realistic seeing value of 0.5 arcsec. For 1m class telescopes, this value seems highly enhanced, approaching a $50\%$ coupling efficiency under the same conditions for almost the entire seeing range studied. This simulation was performed under the assumption of a filling factor of the microlens of $\pi/4$ of the total light.

As can be seen, in the diffraction limit, the coupling efficiency is better when no microlens array is placed. This is because some light is lost due to an imperfect filling factor of the fibre optics arrangement. However, in the presence of atmospheric turbulence, the coupling efficiency improves and becomes more stable if a microlens array is introduced. This stability is related to the size of the microlens array. On the other hand, the complexity of the design, the number of components, and the power consumed by the proposed wavefront sensor increase with the size of the microlens array, and the amount of light gathered by each microlens decreases. Therefore, there is a trade-off between the accuracy of the wavefront sensing, the coupling efficiency of each slice of the wavefront, and the sensitivity of the photodetectors present in the photonic integrated circuit.

\section{Simulation method}

\label{Simulation_method}

In order to verify the approach presented above, in this subsection a general simulation of wavefront sensing and wavefront reconstruction has been performed using the open source tool OOPAO, \citep{heritier:hal-04402878}, for a punctual source.

The system is simulated as a closed-loop AO system. The AO simulation is considered monochromatic. The number of actuators of the DM is considered to be equal to  the number of pixels or microlenses used on the wavefront sensor. The DM is considered a continuous-surface mirror whose actuators follow a Fried geometry. No latency between the detection and the correction of the wavefront is considered in the simulation.

To perform the control loop, an integral controller was implemented in the simulations \citep{s23229186}. Firstly, the deformable mirror is set to be flat. A first measurement of the wavefront sensor is performed, and the shape of the deformable mirror is updated following the law proposed in Eq. \eqref{DM_shape},

\begin{equation}
u_{k} = u_{k-1}-K_{i}T_{k}
\label{DM_shape}
\end{equation}

\noindent where $u_{k}$ denotes the command applied to the deformable mirror actuator $k$; $T_{k}$ refers to the residual wavefront measured by the WFS; and $K_{i}$ is the gain of the integral controller. For simulations, $K_{i}$ was set to 0.4. This value provided a good balance between loop stability, effective correction, and fast actuation. The results presented in the following section are the average and standard deviation of 500 control loop iterations. The first iterations, which corresponded to transients, were removed.

For some particular cases, where the seeing value was $\geq0.9$ arcsec for the smallest microlens arrays (4x4 and 8x8), the value of $K_{i}$ needed to be lowered to 0.1 to ensure the stability of the control loop. The number of iterations also had to be increased by an order of magnitude, as the response of the AO system was slower and larger transients were produced.

A magnitude 8 natural guide star (NGS) is defined in the J2 band, centred at 1550 nm, the common wavelength in integrated photonics. This star is used for performing the wavefront sensing. In this case, the NGS emulates the pinhole mask used in solar observations.

\begin{figure*}[ht]
    \centering

    \includegraphics[width=0.32\textwidth]{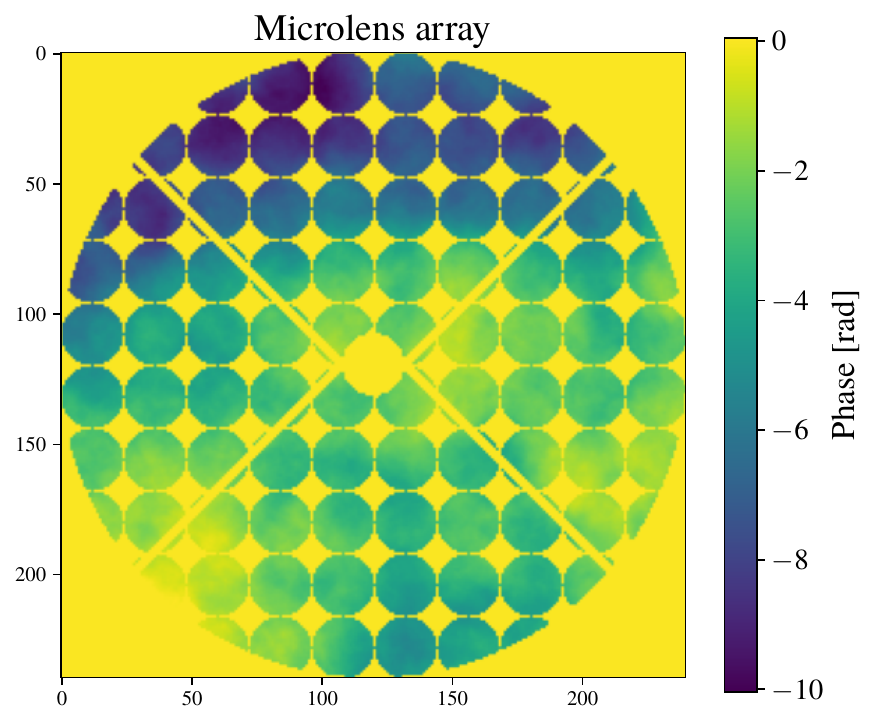}
    \hfill
    \includegraphics[width=0.31\textwidth]{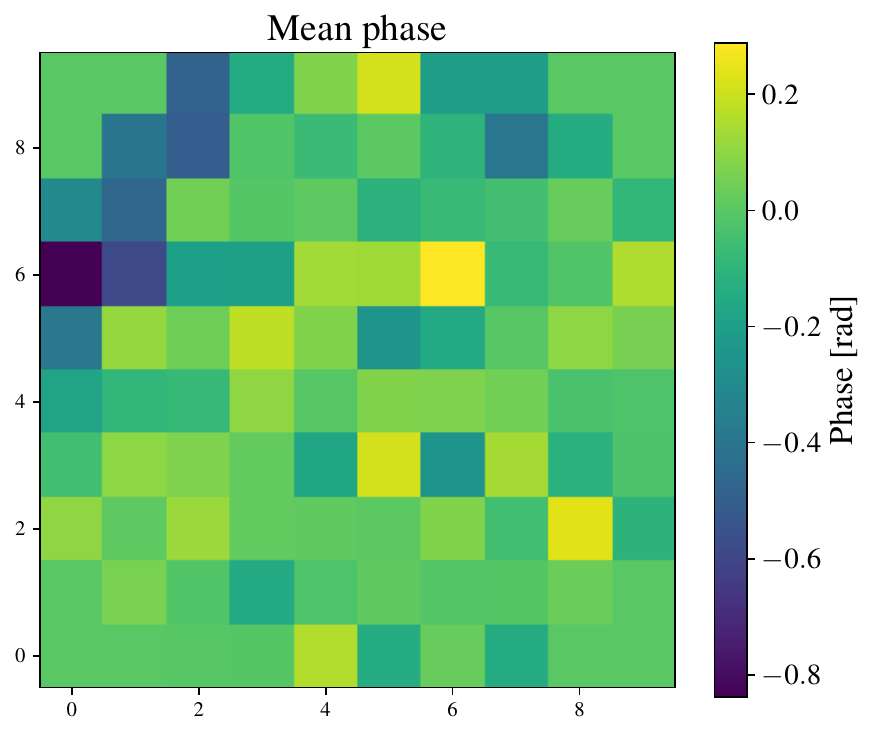}
    \hfill
    \includegraphics[width=0.31\textwidth]{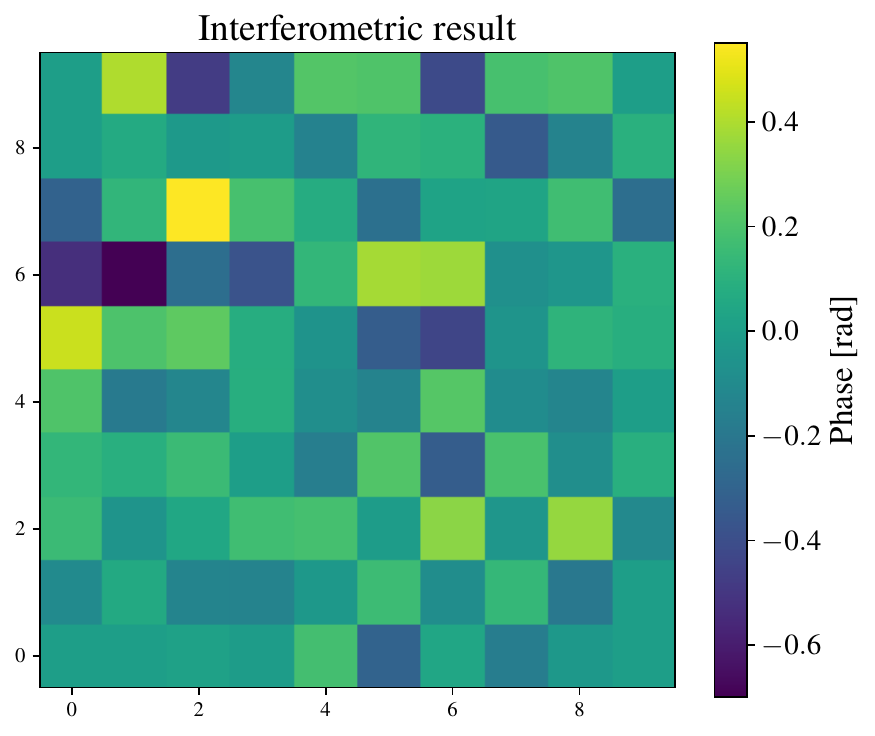}

    \vspace{0.5cm}

    \includegraphics[width=0.33\textwidth]{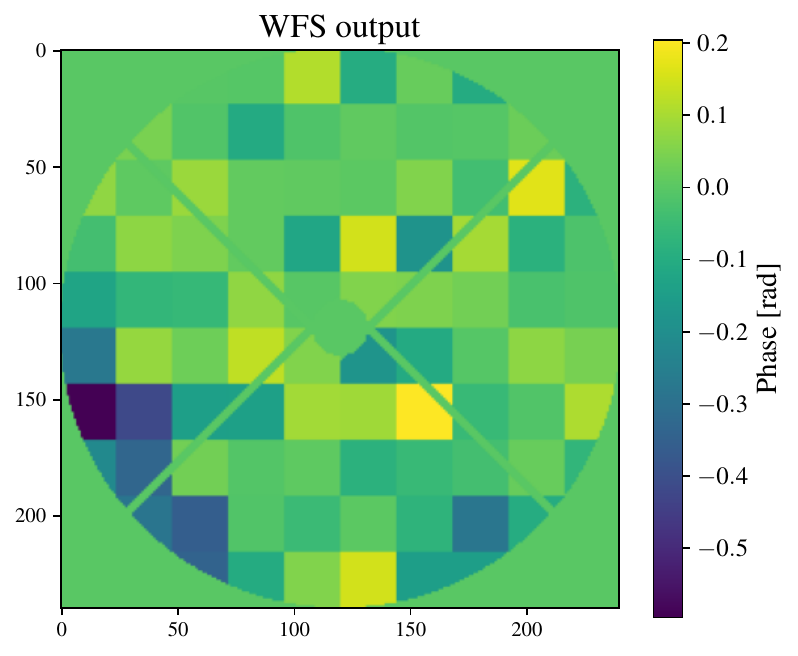}
    \hfill
    \includegraphics[width=0.33\textwidth]{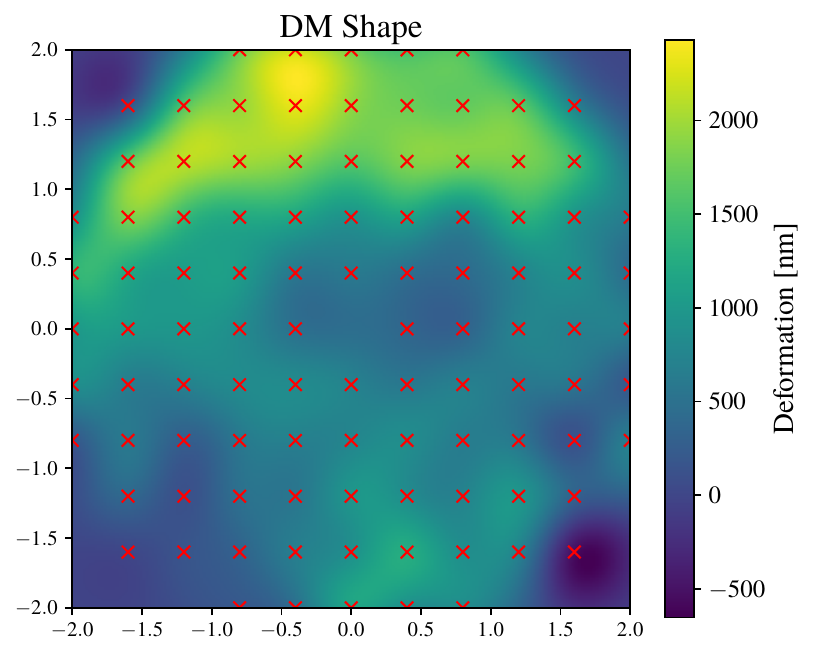}
    \hfill
    \includegraphics[width=0.33\textwidth]{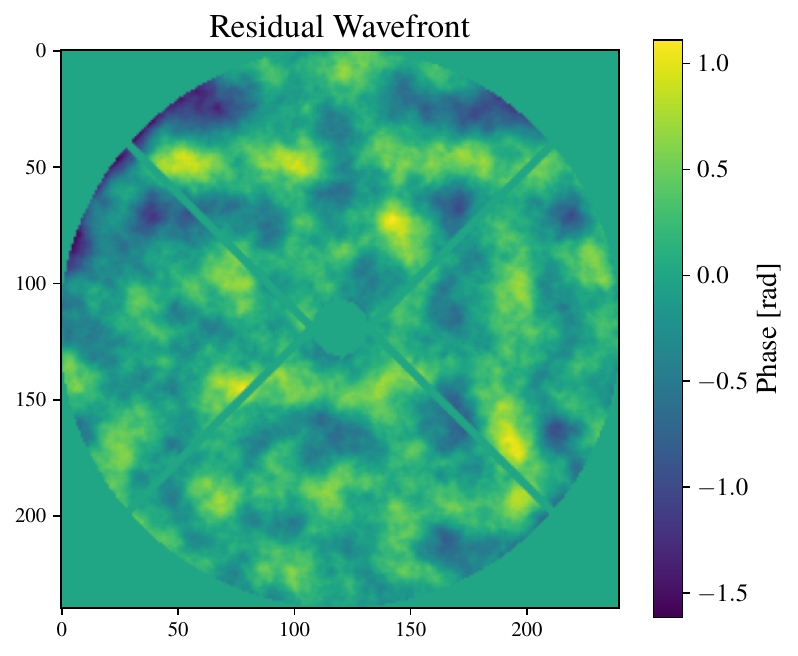}

    \caption{\textit{From top left to bottom right}: Kolmogorov phase screen with overlaid microlens array during the first iteration of the control loop, mean phase per subaperture, interferometric signal at the WFS output, wavefront sensor output, deformable mirror shape, and location of the actuators, marked with a cross, and residual WFE for a 10x10 microlens array at a seeing of 0.5 arcsec after 500 iterations.
    }
    \label{fig:Puil_microlenses}

\end{figure*}

Then, a Kolmogorov phase screen is applied to its wavefront. The defined pupil corresponds to a 4m class telescope, with a central obstruction of 0.1m in diameter and a spider thickness of 0.05m. This simulation was set with a seeing of 0.5 arcsec and a $L_{0}$ of 15 metres. For the sake of reproducibility, the simulation parameters are provided in Table~\ref{Sim_parameters}. Next, an array of microlenses is placed on the pupil plane, as shown in Fig. \ref{fig:Puil_microlenses}. This was calculated by defining a mask that has a zero value outside the pupil mask and a value of one inside. Then, each slice of the pupil image is imaged at the entrance of each optical fibre.

\begin{table*}[h]
\caption{Simulation parameters.}  
\label{Sim_parameters}

\centering                          
\begin{tabular}{c c c c c c|c c}
\hline\hline          
Atmosphere parameters & & & Value & & & Detector parameters & Value\\
\hline  
Altitude [km] & 0 & 1 & 5 & 10 & 20 & Photon noise & True\\               
$Cn^{2}$ profile & 0.45 & 0.1 & 0.1 & 0.25 & 0.1 & Readout noise & 0\\
Wind speed [m/s] & 10 & 12 & 11 & 15 & 20 & Quantum efficiency & 1\\
Wind direction [$\degree$] & 0 & 72 & 144 & 216 & 288 & PSF sampling & Shannon sampl.\\
Seeing [arcsec] & & & 0.5 & & & Binning & 1\\ 
Outer scale (L0) [m] & & & 15 & & & & \\

\end{tabular}

\vspace{-0.11em}

\begin{tabular}{c c| c c| c c}
\hline\hline
Telescope parameters & Value & NGS parameters & Value & STO parameters & Value\\
\hline
Diameter [m] & 4 & Optical band & J2 & Optical band & K\\
Sampling Time [s] & 1/1000 & Central wavelength [nm] & 1550 & Central wavelength [nm] & 2179\\
Central obstruction [m] & 0.1 & Bandwidth [nm] & 260 & Bandwidth [nm] & 410 \\
fov [arcsec] & 10 & Magnitude & 8 & Magnitude & 8\\
Spiders thickness [m] & 0.05 & Coordinates & [0,0] & Coordinates & [1,0]\\ 

\hline
\end{tabular}
\tablefoot{NGS: natural guided star, STO: scientific target object.}
\end{table*}

It is assumed that the phase at the output of each fibre is the average phase across its input. Thus, only a single phase is present at the output of each fibre of the fibre array, resulting in a discretised version of the wavefront. Therefore, the accuracy of the wavefront reconstruction depends directly on the size of the microlens array and the number of microlenses that are placed.

By applying the equation derived in Eq. \eqref{eq:signal_description} and using the snake pattern presented in Fig. \ref{fig:phase_measurements}, an interferometric pattern that represents the phase difference between each optical fibre is obtained. 

It is assumed that the intensity across the pupil is constant, so intensity variations are not considered. In the practical case where intensity variations are introduced by differences in the coupling efficiency between optical fibres, with the photodetectors $PD_{2N}$ in Fig. \ref{fig:ipwfs_circuits}, the intensity at each of the inputs can be known. Using the information from the interferometry and knowing the intensity at each of the inputs, the phase difference can be calculated.

This interferometric pattern represents the phase difference between two adjacent pixels. This would be the output received from the IP-WFS. A reconstruction algorithm should then use this data to reconstruct the wavefront shape, and therefore the shape of the deformable mirror. For this algorithm, the data obtained from the interferometry are going to be stacked in a matrix following a snake pattern. 

As explained earlier, to obtain the sign of the wavefront slice, a second measurement must be performed (see Fig. \ref{Calibration_explaination}).  For this second measurement, the phase modulators present in the circuit are considered to introduce a small phase shift of $\zeta = \theta_{1}(I_{\rm dc2}) - \theta_{2}(I_{\rm dc4}) = 1~\text{mrad}$. Then, by subtracting the values obtained for the two measurements and taking the sign function of the resulting value, the sign can be calculated. The phase then can be retrieved using Eq. \eqref{Two_measurements}:

\begin{equation}
\Delta \phi_i
= \operatorname{sgn}\Bigg(
I_{p,i}(\mathbf{r}, 0) - I_{p,i}(\mathbf{r},\zeta)\Bigg)\arccos\Bigg(\frac{I_{p,i}(\mathbf{r},0)}{2I(\mathbf{r})}-1
\Bigg).
\label{Two_measurements}
\end{equation}

\noindent Here $\Delta \phi_i=\phi_{i+1}-\phi_{i}$, $I_{p}(\mathbf{r},0)$ represents the first measurement, without any phase delay, and $I_{p}(\mathbf{r}, \zeta)$ represents the second measurement with a phase delay $\zeta$. These intensity measurements are calculated using Eq. \eqref{eq:signal_description}.

The final interferometric pattern, presented in Fig. \ref{fig:Puil_microlenses}, is then obtained. The phase shifts that should be applied by each actuator in the DM are denoted by $T_{k}$ in Eq. \eqref{DM_matrix},

\begin{equation}
    T_{k}=\sum_{i=1}^{k}\Delta \phi_i
\label{DM_matrix},
\end{equation}

\noindent where the subindex ${k}$ corresponds to each of the actuators of the DM and follows the snake pattern. As can be seen, the phase information at each mirror depends on the phase information of the previous one. This is the reason why the configuration presented in Fig. \ref{fig:ipwfs_circuits} will fail in the case that one pixel of the WFS gets damaged, resulting in a systematic error.

After these mathematical transformations, the resultant phase shift that should be applied to each mirror is obtained. This phase corresponds to the negative of the phase present at the output of the optical fibres, with a certain phase offset, which is negligible, since it does not affect the flatness of the wavefront.

These results show the reconstruction of the discretised wavefront presented in Fig. \ref{fig:Puil_microlenses}, verifying the proposed reconstruction algorithm. It can be observed that some small-scale structures that correspond to high spatial frequencies of the wavefront could not be corrected, mainly because the size of the structures that can be corrected is limited by the size of the microlens array. If a higher level of correction is to be achieved, a larger number of pixels in the wavefront sensor and actuators in the deformable mirror is needed.

Since the WFS needs to perform two measurements, one to estimate the absolute value of the phase differences between the wavefront slices and another to determine their sign, the AO loop will be effectively running at half of its sampling time. In the particular case of this simulation, the AO system will run effectively at $500~\rm Hz$.

As phase errors are inversely proportional with $\lambda$, if the correction and sensing wavelengths differ, a correction factor must be applied. This correction factor would be the ratio of the target to the sensing wavelengths, $\lambda_{STO}/\lambda_{WFS}$, to the phase shift of each mirror. This is an approximation that can be valid when the values of $\lambda_{STO}$ and $\lambda_{WFS}$ are similar. If this is not the case, a more in-depth analysis of the effects of measuring and correcting at different wavelengths is required.

\section{Results}

\label{results}

The corrected wavefront obtained through the simulation method described in Sect. \ref{Simulation_method} is illustrated in Fig. \ref{fig:Puil_microlenses}. This correction was achieved by applying the wavefront generated at 1550 nm to the wavefront of a target source in the K-band with a seeing of 0.5 arcsec. Some scientific cases in this optical band are the study of the coolest regions of sunspots located at their umbra using the Ti\,{\sc i} spectral lines at 2200 nm \citep{Smitha} or the probe of the magnetic and velocity fields in the sunspot umbra and penumbra at 2231 nm \citep{Livingston2003}.

\begin{figure}[h]
    \centering
    \includegraphics[width=0.2435\textwidth]{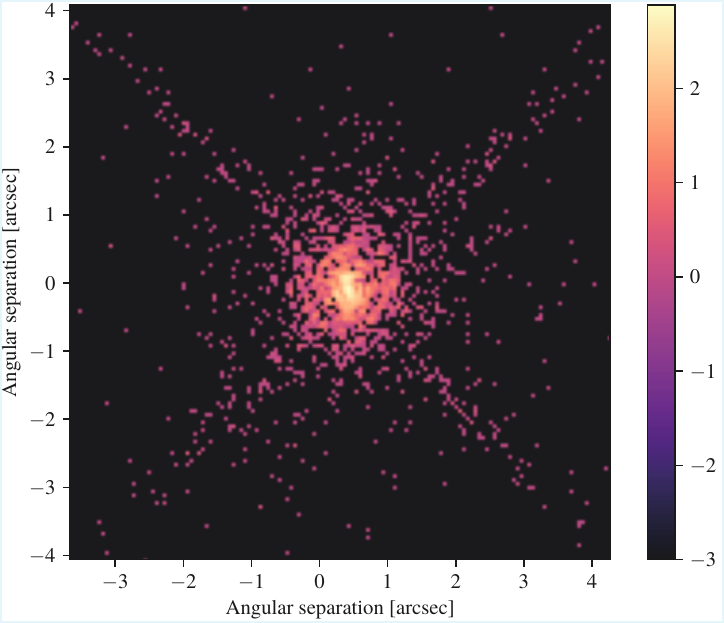}
    \hfill
    \includegraphics[width=0.2405\textwidth]{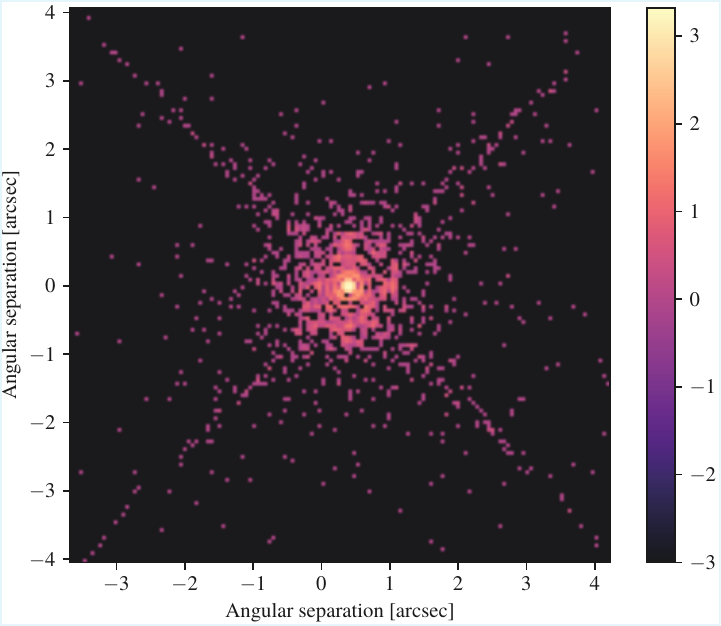}
    \caption{PSF before and after the wavefront correction in the logarithmic scale. The wavefront sensing was performed at a wavelength of 1550 nm using a natural guide star, and the correction was applied to a point source in the K-band.}
    \label{fig:PSF_improvement}
\end{figure}

As can be seen, the corrected wavefront is more uniform compared to the uncorrected version. The PSFs obtained are shown on a logarithmic scale in Fig. \ref{fig:PSF_improvement}, where a non-ideal science camera was used in the simulation. After applying the wavefront correction, the PSF exhibits sharper and more concentrated behaviour, with a brighter core and reduced halo caused by light dispersion in the atmosphere.

\begin{figure}[h!]
    \centering
    \includegraphics[width=0.45\textwidth]{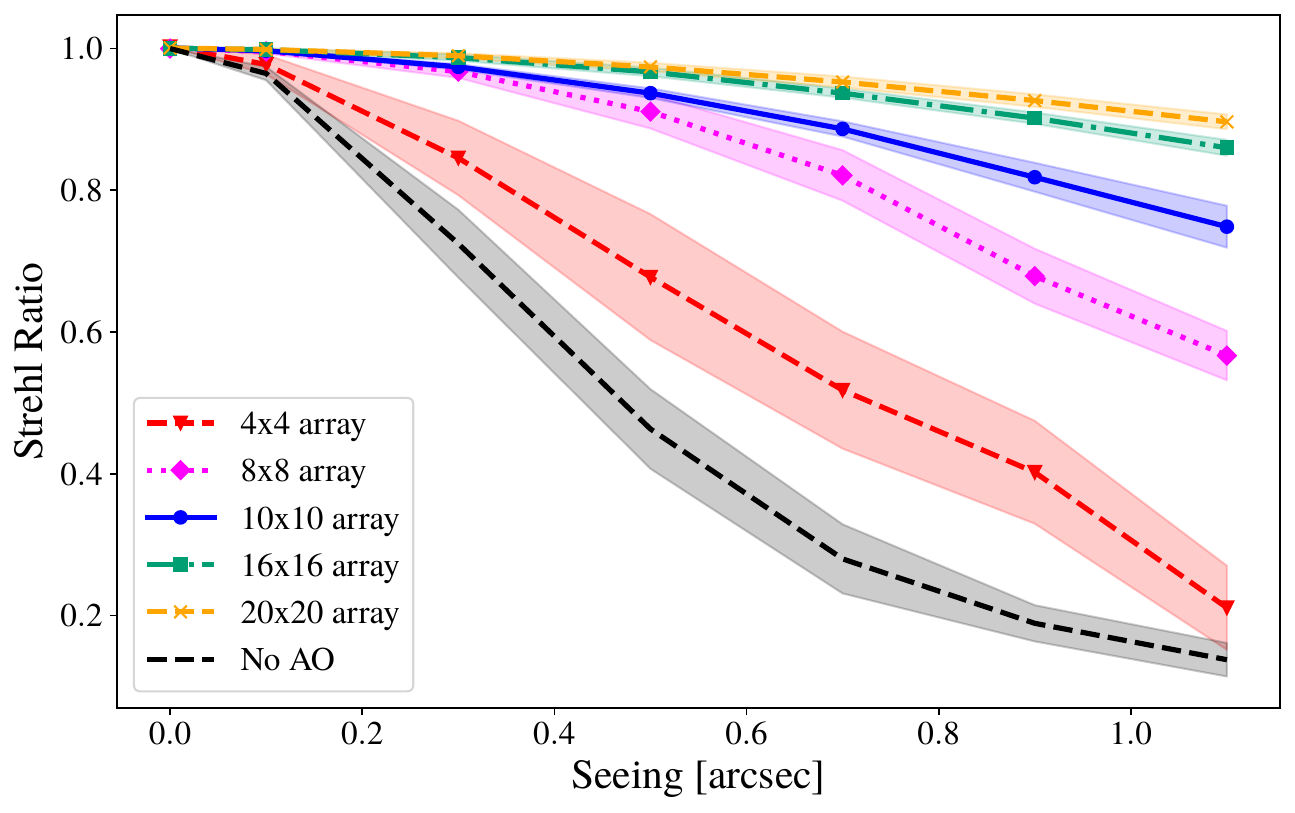}
    \hfill
    \includegraphics[width=0.45\textwidth]{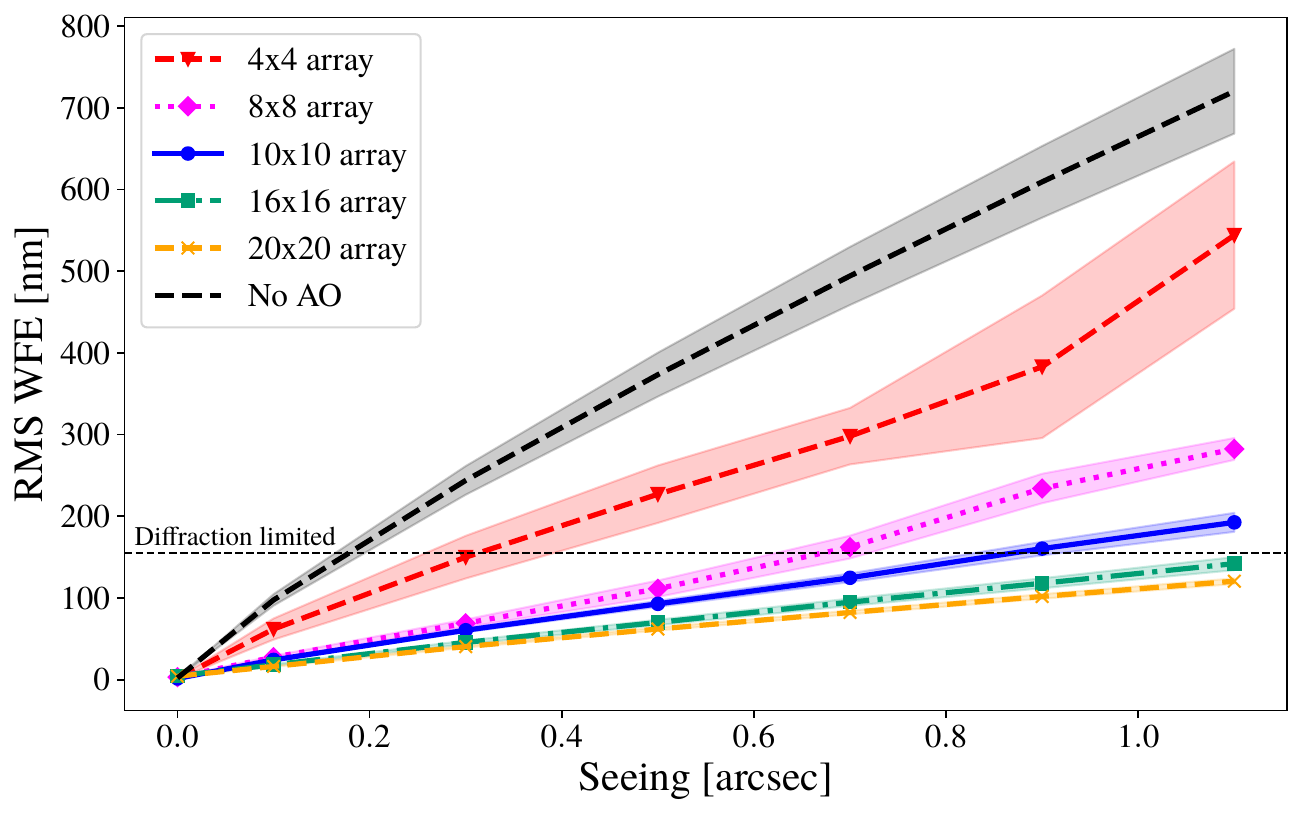}
    \hfill
    \includegraphics[width=0.45\textwidth]{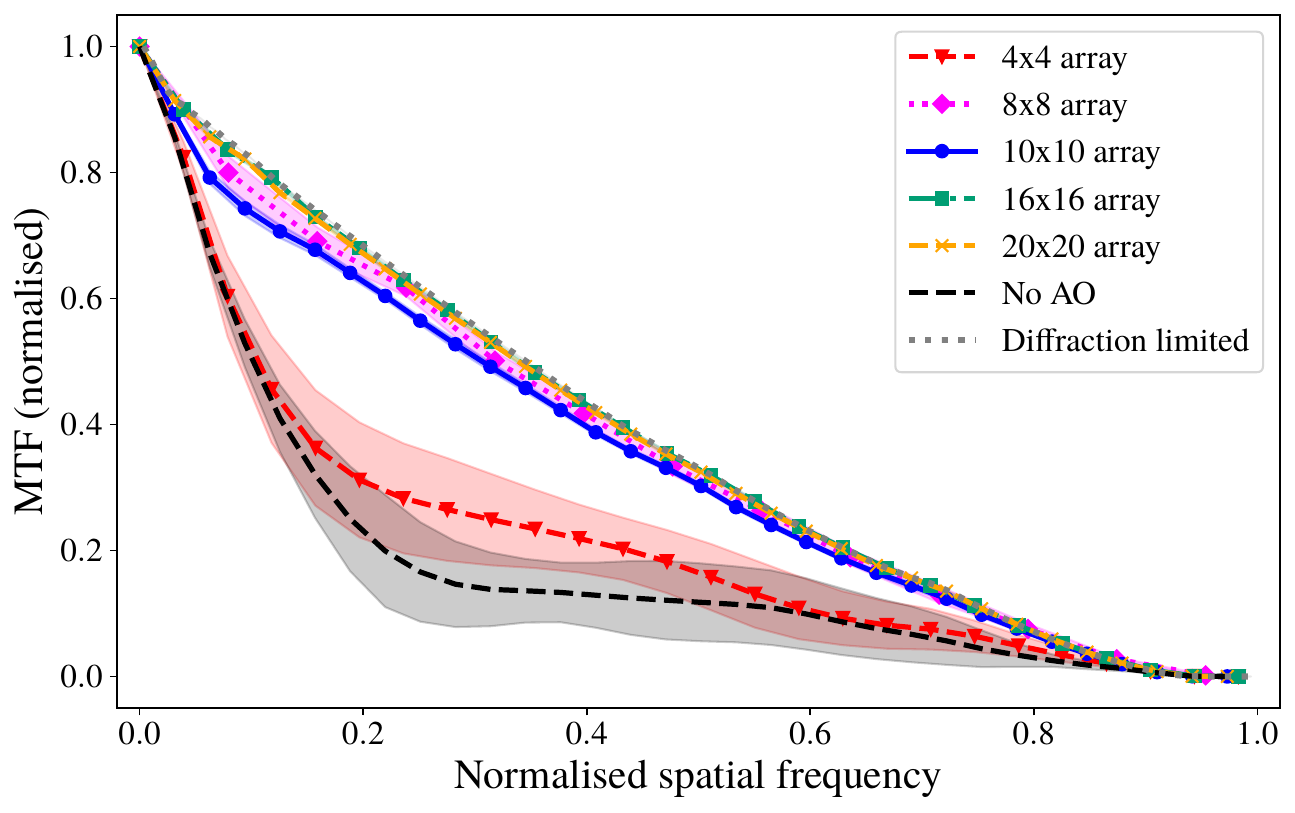}
    \hfill
    \includegraphics[width=0.45\textwidth]{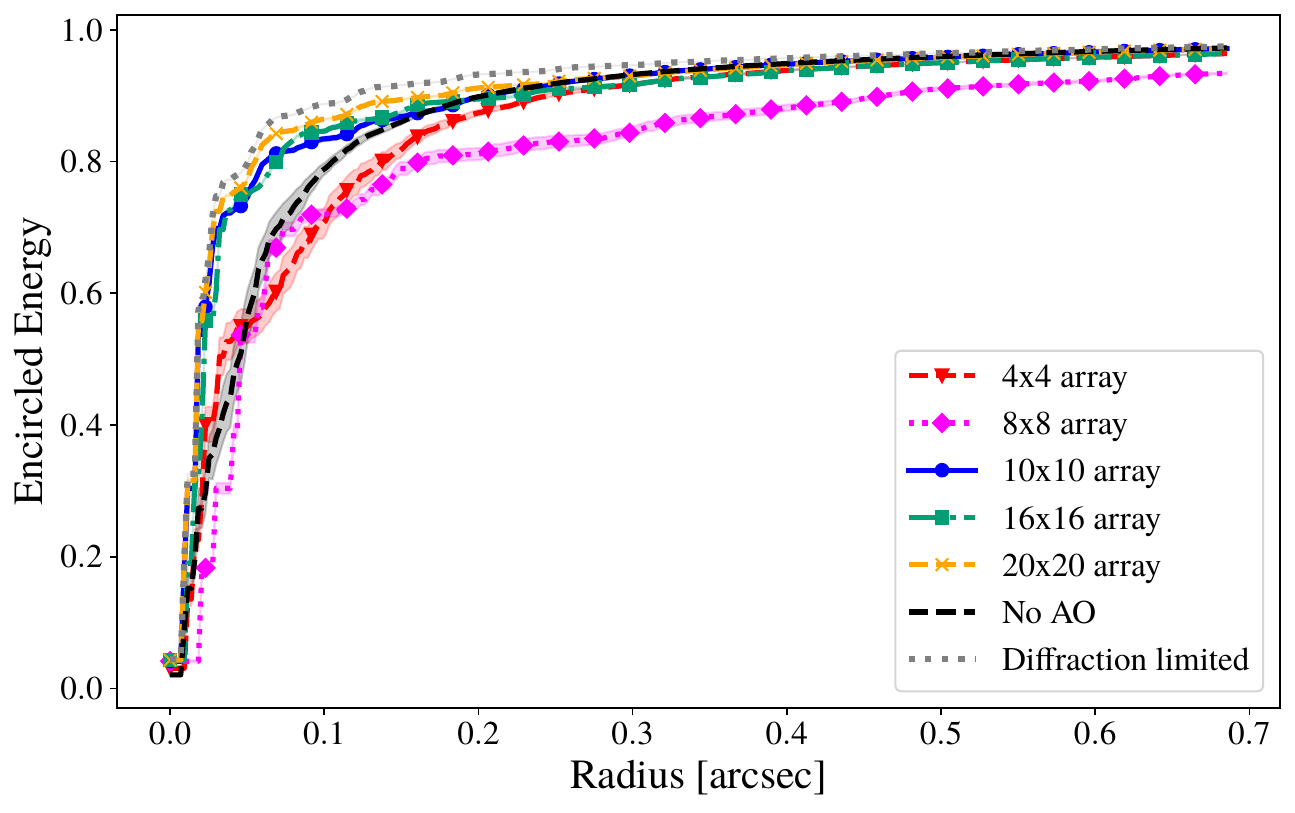}
    \caption{SR (\textit{top}) and WFE (\textit{top middle}) as a function of seeing for different microlens arrays. \textit{Bottom middle}: Normalised MTF at a seeing value of 0.5 arcsec with a spatial cutoff frequency of $1836/\text{f}$ cycles/mm. \textit{Bottom}: EE at a seeing of 0.5 arcsec. Simulations were performed for short-exposure frames of 1 ms on a 4m telescope.}
    \label{fig:SR_WFE_4m}
\end{figure}

Figure \ref{fig:SR_WFE_4m} depicts the simulation of different adaptive optics system performance metrics, including the Strehl ratio (SR), the root mean square (RMS) WFE, the modulation transfer function (MTF) and the encircled energy (EE) for different microlens arrays for a 4m class telescope.

The SR is computed by comparing the peak intensity of the point spread function in the absence of atmospheric turbulence against the peak intensity when turbulence is considered in the model. The calculation is performed for short exposures, assuming a camera integration time of $1~\rm ms$.

The SR improves when an array of microlenses is introduced. For realistic seeing conditions of 0.5 arcsec, the SR improves from $0.463\pm0.056$ to $0.974\pm0.006$, representing an improvement of $51.1\%$ for a 20x20 array, with a maximum increase value of $75.87\%$ achieved at a seeing of 1.1 arcsec. However, this SR could be further improved by introducing larger microlens arrays, with the drawback of gathering a smaller amount of light per lens. This improvement is mainly driven by the reduction of the fitting error between the original wavefront and the reconstructed version due to its discretisation.

The RMS WFE measures the deviation of the measured wavefront from a flat one. This value has been calculated using Eq. \eqref{WFE_eq} for a discrete set of data points,

\begin{equation}
\mathrm{WFE} = \sqrt{\frac{1}{n}\sum_{i=1}^{n} \left(\Delta x_i\right)^2},
\label{WFE_eq}
\end{equation}

\noindent where n represents the number of pixels across the pupil and $x_{i}$ is the value of each pixel.

The results show how the WFE decreases as a function of the size of the microlens array. It also shows how a value of $\lambda/14$, which is commonly defined as the limit where an optical system can be considered diffraction limited \citep{Kahil} (dashed line in Fig. \ref{fig:SR_WFE_4m}), is achieved for the whole seeing range studied for 16x16 and 20x20 microlens arrays. For an array of 10x10 microlens, the system can be considered diffraction limited for seeings < 0.9 arcsec. For 8x8 and 4x4 arrays, they can be considered  diffraction limited at seeings > 0.7 and > 0.3, respectively.

The MTF is a parameter that measures the ability of an AO system to reproduce the contrast and resolution of the observed target in the final image. This is of significant importance in solar observations, as the solar surface contains fine, low-contrast features that need high MTF values to be resolved.

It is defined as the modulus of the Fourier transform of the PSF. The spatial cutoff frequency, defined as $f_{0}=\frac{D}{\lambda f}$, has a value of $1836/\text{f}$ cycles/mm, where f stands for the focal length of the telescope.

The results show how the system can recover the contrast at spatial frequency that was deteriorated by the atmospheric turbulence at a seeing value of 0.5 arcsec, getting a response close to a diffraction-limited system for arrays $\geq$ 8x8 microlenses. This is primarily due to the fact that when larger microlens arrays are employed, smaller sections of the wavefront are analysed, thus enabling smaller wavefront structures to be detected and corrected. Consequently, the spatial frequency response of the AO system is improved.

The 4x4 microlens arrays show a small improvement in the results for mid-spatial frequencies, between 0.2 and 0.6, with a maximum improvement of 11.41$\%$ at a normalised spatial frequency of 0.275. Instead,  little to no enhancement is observed at low and high spatial frequencies, where the maximum improvements observed are below 7.53$\%$ and 1.72$\%$, respectively, thus resulting in a performance level that approaches that of the uncorrected system.

Increasing sampling to larger microlens arrays leads to an improvement in mid-spatial frequencies compared to the 4x4 configuration, with a maximum enhancement  of 44.8$\%$ at a spatial frequency of 0.22 for a 20x20 microlens array, along with an improvement in low spatial frequencies, between 0.05 and 0.2 up to 46.56$\%$ with respect to the uncorrected system. At high-spatial frequencies an improvement below 11.59$\%$, is observed. At very high frequencies, the uncorrected system approaches the diffraction-limited response.

The last performance metric studied is the EE. This parameter measures how much of the total energy is contained within a given region defined by a circle of radius R. The plot shows the results for different radius sizes, ranging from 0 to 0.6 arcsec, in the diffraction limit, without AO and for different microlens array configurations, at a seeing value of 0.5 arcsec.

The results show how, by using larger microlens arrays, the response of the EE approaches the diffraction limited performance. In particular, the radius containing the $50\%$ of the total energy of the PSF (EE50) shows a reduction from 0.0445 to 0.0178 arcsec when adaptive optics is implemented with a 20x20 microlens array. This represents an improvement, which brings the system closer to the diffraction limit, which is found at 0.0178 arcsec.

The study finds a little improvement in the results by using a 4x4 microlens array on an AO closed-loop system. These results are highly improved for larger microlens arrays across the entire range of studied seeing values, as well as an improvement of the spatial frequency response of the system.

In practice, the performance metrics studied during this analysis are expected to degrade, since only atmospheric distortion of the wavefront has been considered for the analysis. In a real scenario, additional sources of wavefront distortion will be present. These include imperfections and aberrations in all optical and photonic elements involved in this wavefront sensor. Furthermore, the optics present in the telescope and in the AO system will contribute to  additional distortion.

\section{Summary}

\label{Summary}

This work presents and analyses the implementation of an integrated-photonic-based wavefront sensor. This integration provides a cost-effective low-power alternative to traditional wavefront sensors that is also smaller in size.

The integration of photonic elements on-chip is relatively recent development. The challenges that this industry needs to address in the coming years, such as high-density integration of photonic components, high-density compatibility with CMOS process, or loss reduction, should be noted. 

Simulations presented in terms of coupling efficiency show the feasibility of coupling light to this device by using an array of optical fibres or  grating couplers. A method for efficient injection of light into this particular instrument using a microlens array has been presented. 

A mathematical analysis then shows the relationship between the amplitude and phase of each of the inputs and the electrical power received by the photodetector. This analysis provides a theoretical background of the performance of the wavefront sensor. Finally, a complete simulation of all the steps that the system should follow to correct wavefront aberrations is presented, with results that verify the feasibility of wavefront correction using this technique. 

Integrated photonics is a novel and promising technology with the potential to enable the development of novel astronomical instruments. Although some of the features that this technology currently presents in terms of working wavelengths, efficiency, sensitivity, and integration of components for high-resolution instruments may not seem sufficient, this technology is improving rapidly, often in response to the requirements of specific applications. Consequently, new opportunities and applications of integrated photonics for a wide range of instruments are emerging, with the potential to improve scientific results.

The direct application of this novel technology to solar adaptive optics could lead to the development of future instruments that could help scientists study and explain novel physical phenomena of the Sun. The work presented enables the possibility of performing wavefront sensing with no need to form images. The results demonstrate how this wavefront sensing could be used in SCAO systems, and how the SR could be improved using the methodology proposed. This has been demonstrated for telescope sizes that range from 1m to the current largest 4m solar telescopes.

The results presented in this work show, using performance metrics such as the SR, the RMS WFE, the MTF, and the EE, how the quality of the final image could be improved with this new wavefront sensing method for microlens arrays > 2x2, for seeing values < 1.1 arcsec.

Future work will include the selection of an adequate photonic design platform, prototype design, laboratory testing of this wavefront sensor on a solar adaptive optics bench, and experimental analysis of the coupling efficiency and the proposed wavefront sensing methodology.

\begin{acknowledgements}
      The authors would like to thank the support from the State Research Agency (AEI) of the Spanish Ministry of Science and Innovation (MCINN), Canary Islands Government and the European Regional Development Fund (FEDER) under grant with reference EQC2019-006594-P, EMIAC project; the \guillemotleft Hub Nacional de Excelencia en Comunicaciones Cuánticas\guillemotright, the Ministerio para la Transformación Digital y de la Función Pública, the European Union–NextGenerationEU; and the Plan de Recuperación, Transformación y Resiliencia. D.P.R. is "personal contratado predoctoral del Programa de Astrofísicos Residentes del IAC".
\end{acknowledgements}

   \bibliographystyle{aa} 
   \bibliography{references}

\end{document}